## CSP-free adaptive Kriging surrogate model method for reliability analysis with small failure probability

Wenxiong Li*, Rong Geng, Suiyin Chen
College of Water Conservancy and Civil Engineering, South China Agricultural University, Guangzhou 510642, China
*Corresponding author. E-mail: leewenxiong@scau.edu.cn

**Abstract** In the field of reliability engineering, the active learning reliability method that amalgamates Kriging model and Monte Carlo simulation has been devised and proven to be efficacious in reliability analysis. Nevertheless, the performance of this method is sensitive to the magnitude of candidate sample pool, particularly for systems with small failure probabilities. To surmount these limitations, this paper proposes an active learning method that obviates the need for candidate sample pools. The proposed method comprises two stages: construction of surrogate model and Monte Carlo simulation for failure probability estimation. During the surrogate model construction stage, the surrogate model is iteratively refined based on the representative samples selected by solving the optimization problem facilitated by the particle swarm optimization algorithm. To achieve an optimal balance between solution accuracy and efficiency, the penalty intensity control and the density control for the experimental design points are incorporated to modify the objective function in optimization. The performance of the proposed method is evaluated using numerical examples, and results indicate that by leveraging an optimization algorithm to select representative samples, the proposed method overcomes the limitations of traditional active learning methods based on candidate sample pool and exhibits exceptional performance in addressing small failure probabilities.

## 1. Introduction

Engineering systems are subject to a multitude of uncertain factors that can significantly impact their reliability and safety. Consequently, reliability analysis has garnered considerable attention in diverse engineering domains such as mechanical systems, infrastructure engineering, automotive, and industrial robotics.

Over the course of several decades, a multitude of reliability methods have been introduced.

Gradient-based methods such as the First Order Reliability Method (FORM) and Second Order Reliability Method (SORM) exhibit low computational complexity but impose restrictive regularity conditions on the Limit-State Function (LSF) that delineates the failure domain. FORM and SORM are classical approaches to conducting reliability analysis with single limit state. They entail first-order and second-order Taylor expansions of the LSF, with the Most Probable Points (MPP) and value of the reliability index determined iteratively. However, due to the omission of high-order terms in the expansion, the solution precision of FORM and SORM is relatively low, rendering them unsuitable for addressing problems with highly nonlinear LSFs. More seriously, FORM and SORM are difficult to achieve ideal results in system reliability analyses. In contrast, sampling techniques based on the variants of the Monte Carlo Simulation (MCS) [1] relax the restrictive regularity conditions at the expense of elevated computational complexity, compared to FORM and SORM. They ascertain the failure probabilities of components or systems through random simulation and statistical testing and have ability to address a wide range of reliability problems with highly nonlinear and complex LSFs. Generally, sampling techniques can be bifurcated into two categories: standard MCS (referred to as MCS in this paper) and Importance Sampling (IS) [2]. Standard MCS typically necessitates a large number of random samples and numerical simulations (such as finite element analysis) to determine the LSF value for each sample. This results in a significant time cost for reliability analysis. Consequently, standard MCS is often utilized as a benchmark to gauge the merits and demerits of other methods but is challenging to directly apply to engineering works. IS can be perceived as an enhanced version of MCS that mitigates the computational complexity of MCS while preserving its mathematical flexibility. In general, two classes of IS methods can be identified [2]. The first class of IS develops the IS density by minimizing a discrepancy measure between the optimal solution and a family of approximating densities. For instance, Au and Beck [1] proposed an efficient sampling algorithm based on a Markov Chain Monte Carlo simulation (MCMC). Jia et al. [3] proposed the density extrapolation approach for efficient estimation of failure probability by integrating density estimation with subset simulation. The second class of IS is based on the general form of the optimal solution in which the computationally demanding LSF is replaced with a fast-to-evaluate surrogate. This class of IS is also referred to as the surrogate model method. Surrogate model methods have emerged as an efficient solution scheme for system reliability analysis in recent years. These methods generally involve two steps: first, constructing a surrogate model to approximate the LSF based on Design of Experiment (DoE) data, which encompasses sample points for system uncertainties and corresponding LSF values; second, performing MCS using the established surrogate model in lieu of the LSF to calculate the system's failure probability. Unlike standard MCS, surrogate model methods can achieve significantly higher efficiency as repeated numerical simulations are required only for the samples in DoE, rather than all samples in MCS. Commonly employed surrogate models include polynomial response surface models [4], radial basis function models [5], Kriging models [6], neural network models [7], support vector machine models [8], and their combination models. The accuracy of the surrogate model is intimately linked to DoE data. Conventional approaches for generating DoE data based on random sampling frequently fail to select representative sample points, and it can be challenging to determine the appropriate scale of DoE. Adaptive surrogate model methods, developed in recent years, have offered an innovative solution for reliability analysis.

The primary differentiation between adaptive surrogate model methods and conventional surrogate model methods lies in the selection of samples for DoE. In the former, the surrogate model is constructed through incremental refinement, with a learning function typically employed to guide the selection of representative samples during the model construction process. In recent years, scholars have extensively debated the implementation and performance of adaptive methods based on various surrogate models, including support vector machine models [9, 10], radial basis function models [11], neural network models [12, 13], and Kriging models [14, 15]. Due to its precise interpolation and effective estimation of uncertainty, the Kriging model is frequently used to establish adaptive surrogate model methods. Echard et al. [16] proposed the Active learning reliability method combining Kriging model and MCS (AK-MCS). Faurat and Gayton [17] introduced a method integrating AK-MCS with efficient global reliability analysis to analyze system reliability with multiple failure modes. Zhang et al. [18] employed the Reliability Expectation Improvement Function (REIF) to develop an active learning surrogate model method. Yun et al. [19] proposed an adaptive Kriging surrogate model method based on a modified U-learning function for system reliability analysis of problems with multiple failure modes. Xiao et al. [15] introduced an adaptive Kriging surrogate model method with a novel learning function based on cross-validation that accounts for both the epistemic uncertainty of surrogate models and the aleatory uncertainty of random variables. Xiao et al. [20] further explored the adaptive Kriging surrogate model method by considering multiple failure modes and mixed variables. Ma et al. [14] conducted a comparative study on the impact of initial DoE data, learning functions, and stopping criteria on the solution performance of an adaptive Kriging surrogate model method. Wang et al. [21] derived a quantificational error measure of Kriging models and presented the modified stepwise accuracy-improvement strategy to solve system reliability problems. Xiong and Sampath [22] proposed three methods including the candidate size control method, multiple trends method and weighted clustering method to improve algorithm performance of AK-MCS and give it the potential for parallel computing. Xiao et al. [23] proposed a learning function for selecting new training samples for complex systems and improved the computational efficiency of system reliability analysis by combining dependent Kriging predictions and parallel learning strategy. Ma et al. [24] proposed a novel method for multiple LSFs within a single run including an adaptive Kriging-Monte Carlo simulation for multiple responses and an adaptive Kriging-generalized subset simulation.

Within the framework of AK-MCS, representative samples required for updating the surrogate model are typically obtained using a selection strategy based on the Candidate Sample Pool (CSP). However, the performance of AK-MCS is sensitive to the size of CSP. Especially, to estimate small failure probabilities, a large-scale CSP is necessary to obtain robust and convergent failure probability estimates. The process of updating the Kriging model based on a large-scale CSP is time-consuming and reduces the efficiency of AK-MCS for estimating small failure probabilities. To enhance the efficiency of AK-MCS, numerous improved AK-MCS methods have been proposed. Echard et al. [25] proposed the method by combining AK-MCS with IS (AK-IS). Balesdent et al. [26] and Zhao et al. [27] conducted research on the combination of an adaptive Kriging surrogate model and IS. Yun et al. [28] proposed the method that combines an adaptive Kriging surrogate model and modified IS to further distinguish important regions from unimportant ones based on AK-IS, with a Kriging surrogate model constructed in the important regions. Sun et al. [29] presented an adaptive Kriging surrogate model method based on the Least Improvement Function (LIF), which combines MCMC to address problems with nonlinear LSFs and small failure probabilities. Building on AK-MCS and K-weighted-means clustering, Lelièvre et al. [30] introduced multipoint enhancement technology and developed an improved AK-MCS for reliability analysis. Lv et al. [31] proposed a novel learning function based on information entropy for the Kriging method and introduced a method combining active learning Kriging model with line sampling. Xiao et al. [15] developed an adaptive Kriging surrogate model method by introducing adaptive IS. Zhou et al. [32] introduced an adaptive Kriging surrogate model method that employs a point selection strategy with a limited region. Xu et al. [33] proposed a modified algorithm that combines AK-MCS and Modified Subset Simulation (AK-MSS) to estimate small failure probabilities. Yun et al. [34] presented an improved AK-MCS method based on adaptive radial-based IS used to reduce the number of candidate points. Liu et al. [35] developed an enhanced AK-MCS by utilizing an efficient CSP reduction strategy. Song et al. [36] proposed a failure boundary exploration and exploitation framework by combining the adaptive Kriging model and sample space partitioning strategy to improve the computational efficiency and avoid memory problems. Wang et al. [37] proposed a novel adaptive Kriging method by combining sampling region scheme and error-based stopping criterion for structural reliability analysis. The improvements to AK-MCS mentioned above generally aim to reduce the size of CSP while ensuring accurate and robust estimation of failure probability. These methods primarily achieve the reduction through advanced sampling techniques or important region selection, thereby improving the efficiency of constructing a Kriging model for the actual LSF. However, various methods have their own limitations in application. For example, the MPP is required by some methods, which presents challenges for these methods in solving problems with multiple MPPs. Additionally, in some methods with subset simulation, generating conditional samples through MCMC and approximating the LSF corresponding to each intermediate failure event are required, which may also lead to a considerable number of model evaluations.

Given the constraints of CSP-based AK-MCS methods, the evolution towards AK-MCS approaches that are independent of CSP is a growing trend. Substantial advancements have been made in recent years in the study of CSP-free adaptive surrogate models that leverage optimization algorithms [11, 38-40]. Jing et al. [39] developed an adaptive surrogate model technique utilizing the Radial Basis Function, where the "potential" MPP for surrogate model updating is procured by resolving a constrained optimization issue via the Genetic Algorithm (denoted as RBF-GA). In their work, the distances between the identified "potential" MPP and the existing DoE are dynamically controlled by a distance constraint. Meng et al. [40] proposed an innovative reliability method that amalgamates MCS and the Kriging surrogate model, where the representative samples are selected using the Particle Swarm Optimization (PSO) algorithm [41, 42] with an Active Weight Learning function (abbreviated as AWL-MCS). In their work, the learning function is constructed based on three factors: the predicted mean value and Kriging variance of the LSF, as well as the joint probability density function of random variables. These accomplishments have propelled the progression of surrogate model methods. Nonetheless, certain facets of the CSP-free methods still necessitate enhancement. In existing optimization-based adaptive surrogate model methods such as RBF-GA and AWL-MCS, the stopping criterion for constructing surrogate model is delineated with the failure probability. Therefore, repeated evaluation of the failure probability is mandated during the construction of the surrogate model. Although the efficiency of estimating failure probability through MCS has been augmented by utilizing the surrogate model in lieu of the true LSF, repeated evaluation of the failure probability remains time-intensive, particularly for problems with small failure probabilities that necessitate large-scale Monte Carlo populations. Furthermore, the formulation of a more judicious objective function for optimization algorithms in the selection of representative samples remains an area for exploration.

To surmount the limitations of CSP-based adaptive Kriging surrogate model methods and existing optimization-based adaptive surrogate model methods, this paper introduces an improved AK-MCS for reliability analysis. In the proposed method, the surrogate model is iteratively refined based on the representative samples selected by solving the optimization problem facilitated by PSO, with similar stopping criterion as AK-MCS. Consequently, the implementation of the proposed method is bifurcated into two stages: construction of the surrogate model and MCS with the established surrogate model for estimating the failure probabilities. This strategy circumvents the prediction of the LSFs for all samples in the Monte Carlo population during the construction of the surrogate model. Based on the U-learning function used in AK-MCS, the new objective function for the optimization problem is defined, incorporating penalty intensity control and density control for experimental design points. This enables an optimal equilibrium between solution accuracy and efficiency to be attained. Ultimately, the performance of the proposed method is assessed using numerical examples.

## 2. AK-MCS

### *2.1. Kriging model*

The Kriging model, a meta-model for nonlinear interpolation, was originally conceived for application in geostatistics. It has found utility in system reliability analysis as a surrogate model, providing an approximation of the relationship between system inputs and outputs. This includes, for example, the correlation between sampling points and the LSF values corresponding to these points. Generally, Kriging model consists of two parts: a parametric linear regression model and a nonparametric stochastic process. The approximate relationship between any experiment $\mathbf{x}=\left(x_1,x_2,\ldots,x_n\right)^{\mathrm{T}}$ and the response $\hat{G}\left(\mathbf{x}\right)$ can be denoted as

$$\hat{G}\left(\mathbf{x}\right)=\mathbf{f}^{\mathrm{T}}\left(\mathbf{x}\right)\boldsymbol{\beta}_r+z\left(\mathbf{x}\right) \tag{1}$$

where $\mathbf{f}^{\mathrm{T}}\left(\mathbf{x}\right)\boldsymbol{\beta}_r$ represents the deterministic part which gives an approximation of the response in mean, $\mathbf{f}^{\mathrm{T}}\left(\mathbf{x}\right)=\left\{f_1\left(\mathbf{x}\right),f_2\left(\mathbf{x}\right),\ldots,f_k\left(\mathbf{x}\right)\right\}$ represents the basis function vector and $\boldsymbol{\beta}_r^{\mathrm{T}}=\left\{\beta_{r1},\beta_{r2},\ldots,\beta_{rk}\right\}$ is the regression coefficient vector. In this paper, ordinary Kriging model is selected which means that $f_i\left(\mathbf{x}\right)=1\ \left(i=1,2,\ldots,k\right)$. In Eq.(1), $z\left(\mathbf{x}\right)$ is a stationary Gaussian process with zero mean $z\left(\mathbf{x}\right)\sim N\left(0,\sigma^2\right)$, and the covariance between two points of space $\mathbf{x}^i$ and $\mathbf{x}^j$ is defined as

$$\mathrm{Cov}\left[z\left(\mathbf{x}^i\right),z\left(\mathbf{x}^j\right)\right]=\sigma^2R\left(\boldsymbol{\theta},\mathbf{x}^i,\mathbf{x}^j\right) \tag{2}$$

where $\sigma^2$ is the process variance, $\boldsymbol{\theta}=\{\theta_1,\theta_2,\ldots,\theta_n\}$ refers to the parameter vector, and $R(\boldsymbol{\theta},\mathbf{x}^i,\mathbf{x}^j)$ represents the correlation function between $\mathbf{x}^i$ and $\mathbf{x}^j$, which is formulated by

$$R(\boldsymbol{\theta},\mathbf{x}^i,\mathbf{x}^j)=\prod_{d=1}^{n}\exp\left[-\theta_d\left(x_d^i-x_d^j\right)^2\right] \tag{3}$$

where $x_d^i$ and $x_d^j$ refers to the $d^{\text{th}}$ component in $\mathbf{x}^i$ and $\mathbf{x}^j$, respectively.

Given the sample set of DoE $\mathbf{S}_{\text{DoE}}=\left[\mathbf{x}^1,\mathbf{x}^2,\ldots,\mathbf{x}^m\right]$ and the corresponding response set $\mathbf{Y}_{\text{DoE}}=\left[G(\mathbf{x}^1),G(\mathbf{x}^2),\ldots,G(\mathbf{x}^m)\right]$ with $m$ being the number of samples in DoE, the scalars $\beta_r$ and $\sigma^2$ are estimated by

$$\hat{\beta}_r=\left(\mathbf{1}^{\text{T}}\mathbf{R}^{-1}\mathbf{1}\right)^{-1}\mathbf{1}^{\text{T}}\mathbf{R}^{-1}\mathbf{Y}_{\text{DoE}} \tag{4}$$

$$\hat{\sigma}^2=\frac{1}{m}\left(\mathbf{Y}_{\text{DoE}}-\hat{\beta}_r\mathbf{1}\right)^{\text{T}}\mathbf{R}^{-1}\left(\mathbf{Y}_{\text{DoE}}-\hat{\beta}_r\mathbf{1}\right) \tag{5}$$

where $\mathbf{R}$ is the correlation matrix with the component $R_{i,j}=R(\boldsymbol{\theta},\mathbf{x}^i,\mathbf{x}^j)$ in which represents the correlation between each pair of points in DoE, and $\mathbf{1}$ refers to the vector filled with 1 of length $m$. $\hat{\beta}_r$ and $\hat{\sigma}^2$ in Eqs. (4) and (5) are related to the correlation parameters $\theta_i$ through the matrix $\mathbf{R}$, then the value of $\boldsymbol{\theta}$ is required to be firstly obtained by using maximum likelihood estimation:

$$\boldsymbol{\theta}=\arg\min_{\boldsymbol{\theta}}\left(\det(\mathbf{R})\right)^{\frac{1}{m}}\hat{\sigma}^2 \tag{6}$$

According to the Gaussian process regression theory, the system response follows the normal distribution as $G(\mathbf{x})\sim N\left(\mu_G(\mathbf{x}),\sigma_G(\mathbf{x})\right)$. Then, based on the Kriging model established according to the given data of DoE, the best linear unbiased predictor of the response $\hat{G}(\mathbf{x})$ at an unknown point $\mathbf{x}$ is shown to be a Gaussian random variate $\hat{G}(\mathbf{x})\sim N\left(\mu_{\hat{G}}(\mathbf{x}),\sigma_{\hat{G}}(\mathbf{x})\right)$ where

$$\mu_{\hat{G}}(\mathbf{x})=\hat{\beta}_r+\mathbf{r}^{\text{T}}(\mathbf{x})\mathbf{R}^{-1}\left(\mathbf{Y}_{\text{DoE}}-\hat{\beta}_r\mathbf{1}\right) \tag{7}$$

$$\sigma_{\hat{G}}^2(\mathbf{x})=\hat{\sigma}^2\left(1-\mathbf{r}^{\text{T}}(\mathbf{x})\mathbf{R}^{-1}\mathbf{r}(\mathbf{x})+u^{\text{T}}(\mathbf{x})\left(\mathbf{1}^{\text{T}}\mathbf{R}^{-1}\mathbf{1}\right)^{-1}u(\mathbf{x})\right) \tag{8}$$

where $\mathbf{r}(\mathbf{x})=\left[R(\boldsymbol{\theta},\mathbf{x},\mathbf{x}^1),R(\boldsymbol{\theta},\mathbf{x},\mathbf{x}^2),\ldots,R(\boldsymbol{\theta},\mathbf{x},\mathbf{x}^m)\right]^{\text{T}}$ and $u(\mathbf{x})=\mathbf{1}^{\text{T}}\mathbf{R}^{-1}\mathbf{r}(\mathbf{x})-1$.

In Kriging model, the predicted mean value at any point $\mathbf{x}^i$ in DoE is consistent with the true response value, namely $\mu_{\hat{G}}(\mathbf{x}^i)=G(\mathbf{x}^i)\,(i=1,2,\ldots,m)$, and the corresponding Kriging variance is null, namely $\sigma_{\hat{G}}^2(\mathbf{x}^i)=0\,(i=1,2,\ldots,m)$. For any point out of DoE, the Kriging variance is not zero, and its value reflects the accuracy of the prediction results at the point. For clarity, **Fig. 1** demonstrates an example of prediction results of Kriging model within the range of [0,10.0], which include $\mu_{\hat{G}}(x)(x \in [0,10.0])$ and $\mu_{\hat{G}} \pm \sigma_{\hat{G}}(x)(x \in [0,10.0])$, based on the given sample data in DoE. The Kriging variance of Kriging model provides a basis for selecting representative experimental design points. Generally, learning functions can be constructed according to Kriging variance and predicted mean value to realize active learning based on the current Kriging model, and finally a Kriging model that sufficiently approximates the relationship between the experiment $\mathbf{x}$ and the response $G(\mathbf{x})$ can be established.

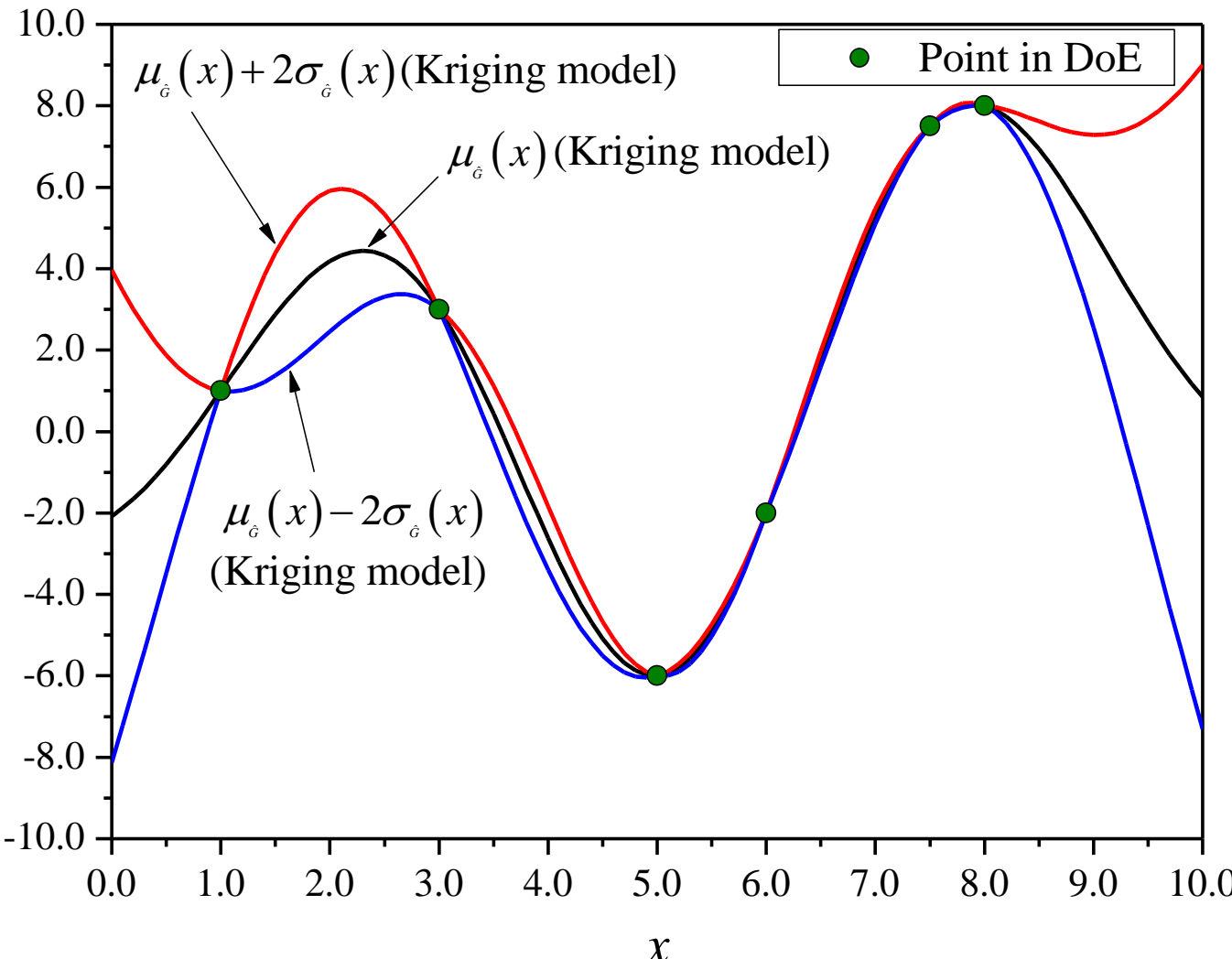

**Fig. 1**. The predicted mean and Kriging variance of Kriging model.

## *2.2. Implementation of AK-MCS*

In AK-MCS [14, 16, 17, 30], a Kriging model is employed to approximate the relationship between sample points and LSF values, and to predict the Limit-State Equation (LSE, $G(\mathbf{x})=0$). Utilizing the constructed Kriging model as a surrogate model allows for the estimation of failure probability via the subsequent formula

$$\hat{P}_f = \frac{n_{\hat{G}<0}}{n_{MC}} \tag{9}$$

where $n_{MC}$ represents the total number of sampling in MCS and $n_{\hat{G}<0}$ refers to number of sampling with $\hat{G}<0$ in MCS.

In order to tackle the challenge of data construction for DoE, researchers have introduced methods utilizing an adaptive Kriging surrogate model with active learning capabilities. These methods involve the progressive refinement of the Kriging model through a sequence of iterative processes. During each iteration, all potential samples are assessed using the current Kriging model. Subsequently, the sample exhibiting the highest (or lowest) value of the learning function is chosen and incorporated into the DoE to update the Kriging model. Typically, representative samples that contribute significantly to the enhancement of the Kriging model exhibit the following spatial distribution characteristics: (1) Representative samples are generally proximate to the LSE; (2) Newly identified representative samples are often situated in areas where existing DoE samples are relatively sparse.

A variety of learning functions have been developed, including the sign indication learning function (U-learning function) [16], the Expected Feasibility Function (EFF) [38], the information entropy theory-based learning function [31], and the LIF [29]. In this study, the sign indication learning function (U-learning function), as utilized in AK-MCS, is employed to guide the selection of representative samples. The U-learning function is constructed based on the risk of crossing the limit state and is defined as follows

$$U(\mathbf{x}) = \frac{\left|\mu_{\hat{G}}(\mathbf{x})\right|}{\sigma_{\hat{G}}(\mathbf{x})} \tag{10}$$

According to the definition of the U-learning function, samples with a high probability of crossing the limit state are incorporated into the DoE set. This is due to the fact that the positive and negative signs of their corresponding LSF values are susceptible to influence from uncertain factors, leading to alterations in failure probability. Consequently, representative samples should be positioned in accordance with the following conditions: (1) The predicted mean value of the LSF is in close proximity to zero; (2) The Kriging variance of the LSF is relatively high. In the process of updating the Kriging model, the values of U-learning function for all candidate samples in CSP $\mathbf{S}_{pool} = \left(\mathbf{x}_c^1, \mathbf{x}_c^2, \ldots, \mathbf{x}_c^p\right)$ are firstly calculated using the current Kriging model, and then the sample with lowest U value among these candidate samples is added to the set of DoE. Hence, the learning criterion is defined as $\min\left(U(\mathbf{x})\right)\left(\mathbf{x} \in \mathbf{S}_{pool}\right)$.

The stopping criterion of Kriging model update process in AK-MCS [16] is set to be $\min\left(U(\mathbf{x})\right) \geq 2.0$, which means a sample is incorrectly classified with a probability of $\Phi(-2) = 0.023$. It is considered that the correct rate of the surrogate model's response signs to the Monte Carlo population can be guaranteed if the stopping criterion is met. Meanwhile, the size of Monte Carlo population should be large enough to reduce the impact of sampling randomness on failure probability estimation. Therefore, the variation coefficient of failure probability $V_{\hat{P}_f}$ expressed as follows should be further accessed

$$V_{\hat{P}_f} = \sqrt{\frac{1-\hat{P}_f}{n_{MC}\hat{P}_f}} \tag{11}$$

When $V_{\hat{P}_f}$ exceeds the preset limit value, it is considered that the size of the current Monte Carlo population is not large enough, and the number of sampling points in Monte Carlo simulation should be increased.

Ref. [16] provides comprehensive details on the implementation of AK-MCS. Given that representative samples are selected from CSP, the distribution of candidate samples significantly impacts the acquisition of representative samples. Generally, the distribution of candidate samples in the peripheral regions of Monte Carlo sampling is relatively sparse, which complicates the procurement of true representative samples for these regions. For problems with small failure probabilities, where true representative sample points often reside at the sampling edge, a large-scale CSP is necessitated to ensure the discovery of true representative sample points. However, as the scale of CSP increases, so does the computational effort required to evaluate candidate samples using the Kriging surrogate model, ultimately affecting the efficiency of adaptive Kriging surrogate model methods. Therefore, overcoming the limitations of CSP-based methods in selecting representative samples is key for adaptive Kriging surrogate model methods to solve reliability analysis problems with small failure probabilities. The authors believe that the utilization of an optimization algorithm to obtain representative samples, rather than the regular CSP-based method, is a promising approach to improving the performance of adaptive Kriging surrogate model methods.

## 3. Improved AK-MCS

By treating the values of random variables as design variables, the selection of representative samples is viewed as an optimization problem. The objective function is typically a nonlinear function with multiple peaks (or valleys). As a result, conventional optimization algorithms, such as gradient-based methods, may only converge to local optima and produce suboptimal results. In contrast, modern optimization algorithms such as genetic algorithm, ant colony algorithm and PSO are more appropriate for solving this complex optimization problem, as they are theoretically capable of finding global optima and do not require the optimized functions to be differentiable or continuous. Given its effectiveness in handling continuous variable optimization and ease of implementation, PSO is employed as the primary algorithm for optimization in the improved AK-MCS. This section introduces the implementation of PSO and the improved CSP-free AK-MCS.

*3.1. Implementation of PSO*

PSO is a global optimization algorithm that was first proposed by Kennedy and Eberhart in 1995 [43]. It is designed to address optimization problems that involve nonlinear objective functions and continuous variables. The underlying concept of PSO is inspired by the foraging behavior of birds, where individuals follow the best-performing member of the flock without knowing the exact location or distance to food. In PSO, each potential solution to the optimization problem is conceptualized as a 'particle' within the search space. Particles are evaluated based on a fitness function determined by the objective function and move through the solution space at random speeds. Through information exchange, particles acquire heuristic information that guides the swarm towards an optimal solution. PSO has found applications in various fields, including structural optimization [44], topology optimization [45], structural reliability evaluation [46] and reliability optimization design [47]. For a more comprehensive understanding and further developments in the application of PSO, readers are referred to Ref. [41].

An optimization problem with a $N_D$-dimension search space can be solved by using PSO with the size of particle swarm (number of particles) $N_{swarm}$ and the maximum iteration number $N_{ite_\max}$ predetermined. When the algorithm runs to the $n^{\text{th}}\left(n=1,2,\ldots,N_{ite_\max}\right)$ iteration, the current position and current velocity of the $i^{\text{th}}\left(i=1,2,\ldots,N_{swarm}\right)$ particle are expressed as $\mathbf{x}_i^{(n)}=\left(x_{i,1}^{(n)},x_{i,2}^{(n)},...,x_{i,N_D}^{(n)}\right)$ and $\mathbf{v}_i^{(n)}=\left(v_{i,1}^{(n)},v_{i,2}^{(n)},...,v_{i,N_D}^{(n)}\right)$, respectively. Then, for each particle, the following formulas are used to update the speed and position in each dimension component at each iteration.

$$v_{i,j}^{(n+1)}=\omega\cdot v_{i,j}^{(n)}+c_1\cdot rand\cdot\left(x_{pbest,i,j}^{(n)}-x_{i,j}^{(n)}\right)+c_2\cdot rand\cdot\left(x_{gbest,j}^{(n)}-x_{i,j}^{(n)}\right)\qquad (j=1,2,\ldots,N_D)\tag{12}$$

$$x_{i,j}^{(n+1)}=x_{i,j}^{(n)}+v_{i,j}^{(n+1)}\qquad (j=1,2,\ldots,N_D)\tag{13}$$

In Eq. (12), $\omega$ is the inertia weight, $c_1$ and $c_2$ refer to the cognition learning factor and social learning factor, respectively, $x_{pbest,i,j}^{(n)}$ represents the $j^{\text{th}}$ component of the historical best position of the $i^{\text{th}}$ particle at the $n^{\text{th}}$ iteration, $x_{gbest,j}^{(n)}$ represents the $j^{\text{th}}$ component of the best position of the whole swarm at the $n^{\text{th}}$ iteration, $rand$ means to generate a random value uniformly distributed within the range of $[0,1]$. For the sake of convenience, the historical best position of a particle and the best position of the whole swarm at the $n^{\text{th}}$ iteration are expressed as $\mathbf{x}_{pbest,i}^{(n)}=\left(x_{pbest,i,1}^{(n)},x_{pbest,i,2}^{(n)},\ldots,x_{pbest,i,N_D}^{(n)}\right)$ and $\mathbf{x}_{gbest}^{(n)}=\left(x_{gbest,1}^{(n)},x_{gbest,2}^{(n)},\ldots,x_{gbest,N_D}^{(n)}\right)$, respectively.

Generally, the historical best position of a particle $\mathbf{x}_{pbest,i}^{(n)}$ changes with the iteration process and can be described by

$$\mathbf{x}_{pbest,i}^{(n+1)} = \begin{cases} \mathbf{x}_i^{(n+1)}, & \text{if } F_{\text{fit}}\left(\mathbf{x}_i^{(n+1)}\right) > F_{\text{fit}}\left(\mathbf{x}_{pbest,i}^{(n)}\right) \\ \mathbf{x}_{pbest,i}^{(n)}, & \text{otherwise} \end{cases} \tag{14}$$

where $F_{\text{fit}}(\bullet)$ refers to the fitness function. Among the historical best positions of all particles, the best position with the highest fitness value is recorded as the best position of the whole swarm, expressed as

$$\begin{aligned} \mathbf{x}_{gbest}^{(n)} &\in \left\{ \mathbf{x}_{pbest,1}^{(n)}, \mathbf{x}_{pbest,2}^{(n)}, \ldots, \mathbf{x}_{pbest,N_{swarm}}^{(n)} \,\middle|\, F_{\text{fit}}\left(\mathbf{x}_{pbest,i}^{(n)}\right) \right\} \\ &= \max\left\{ F_{\text{fit}}\left(\mathbf{x}_{pbest,1}^{(n)}\right), F_{\text{fit}}\left(\mathbf{x}_{pbest,2}^{(n)}\right), \ldots, F_{\text{fit}}\left(\mathbf{x}_{pbest,N_{swarm}}^{(n)}\right) \right\} \end{aligned} \tag{15}$$

As discussed above, the movement of particles is influenced by three factors. The first is the current speed of motion, which reflects the continuity of particle movement between successive iterations. The second is the particle's historical best position, which reflects the influence of its own experience on its direction of movement. The third is the best position of the swarm, which reflects the influence of the swarm's collective experience on the direction of particle movement. **Fig. 2** illustrates the motion of a particle.

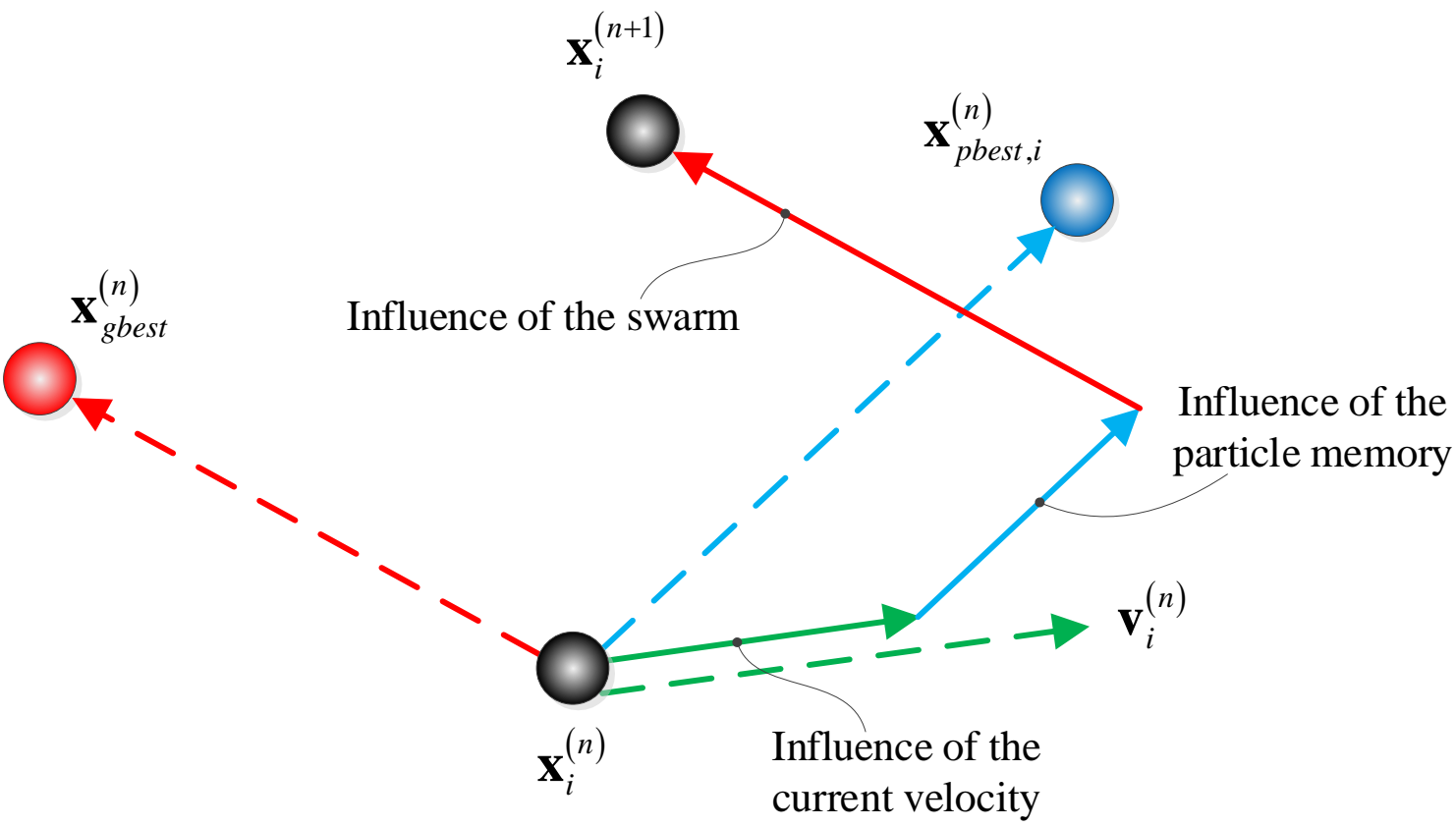

**Fig. 2**. Sketch map of particle motion in PSO [41].

The performance of PSO is influenced by the parameters $\omega$, $c_1$ and $c_2$. The inertia weight $\omega$ is a parameter that controls the influence of a particle's velocity from the previous iteration. Introduced by Eberhart [46], this parameter affects the balance between global and local search capabilities. A higher inertia weight generally enhances global optimization ability, while a lower inertia weight improves local optimization ability. The value of the inertia weight typically ranges between 0 and 1.0, with a recommended value of 0.729 as suggested by Clerc and Kennedy [48]. The cognitive learning factor $c_1$ reflects the influence of the distance between a particle's current position and its historical best position. This means that particles move according to their own cognition. The social learning factor $c_2$ describes the influence of the distance between a particle's current position and the swarm's best position. This reflects information sharing and cooperation among particles, as well as their tendency to move according to the swarm's collective experience. Both $c_1$ and $c_2$ are typically set to 2.0, with a recommended value of 1.494 as suggested by Clerc and Kennedy [48]. Additionally, a velocity limit constant $v_{\max}$ is often introduced to control the maximum distance a particle can move in each iteration, thereby controlling the convergence speed and global search ability of the algorithm.

The main steps of PSO used in this work are as follows.

**Step 1**: Set control parameters for PSO, including $N_{swarm}$, $N_{ite_\max}$, $\omega$, $c_1$, $c_2$ and $v_{\max}$.

**Step 2**: Initialize the particle swarm by randomly generating the position and velocity of each particle within the solution space. Evaluate all particles and set their current position as their historical best position. Determine the swarm's best position according to Eq. (15).

**Step 3**: Update the position and velocity of all particles using Eqs. (12) and (13), and generate a new particle swarm.

**Step 4**: Evaluate all particles in the swarm and determine their historical best position according to Eq. (14). Determine the swarm's best position using Eq. (15).

**Step 5**: If the maximum number of iterations has been reached, output the swarm's best position and end the program. Otherwise, return to **Step 3** and continue.

### *3.2. Implementation of the improved AK-MCS*

#### 3.2.1. Optimization model for representative points

The values of random variables in the standard normal space are typically considered as the design variables and denoted as $\mathbf{u}=\left(u_1,u_2,\ldots,u_{N_D}\right)$. The representative sample is determined by solving the following optimization formulation,

$$\begin{aligned}
&\text{find} \quad \mathbf{u}^*=\left(u_1^*,u_2^*,\ldots,u_{N_D}^*\right) \\
&\min \quad F_{\text{obj}}\left(\mathbf{u}\right)=\frac{\left\|\mu_{\hat{G}}\left(\mathbf{u}\right)\right|-\delta\right|}{\sigma_{\hat{G}}\left(\mathbf{u}\right)}+p\cdot\max\left(\left\|\mathbf{u}\right\|-r_c,0\right) \\
&\qquad \text{s.t.} \quad u_j\in\left[-u_{\lim},u_{\lim}\right]\left(j=1,2,\ldots,N_D\right)
\end{aligned} \tag{16}$$

where $\mathbf{u}^*$ is the optimum of Eq. (16) where $u_j \in [-u_{\lim}, u_{\lim}] (j = 1, 2, \ldots, N_D)$ reflects the search range of the $j^{\text{th}}$ design variable, $\delta$ is a preset small quantity that describes the offset between the search target and the LSE. Given that the probability of obtaining samples in $(-\infty, -6.0)$ and $(6.0, \infty)$ is extremely small when conducting random sampling according to the standard normal distribution, this paper sets $u_{\lim} = 6.0$. In the implementation of PSO, the fitness function can be obtained by $F_{\text{fit}}(\mathbf{u}) = 1/(F_{\text{obj}}(\mathbf{u}) + \delta_0)$ with $\delta_0$ the preset small quantity such as $1.0\times10^{-8}$.

In Eq. (16), $F_{\text{obj}}(\mathbf{u})$ represents the objective function, where $p \cdot \max(\|\mathbf{u}\| - r_c, 0)$ is a penalty term with $r_c$ controlling the sampling range and $p$ representing the penalty intensity, named as penalty coefficient. The purpose of introducing the penalty term is to minimize the likelihood of selecting representative sample points outside the specified range, as these points contribute minimally to the construction of the surrogate model. As illustrated in **Fig. 3**, the boundary region of Monte Carlo sampling forms a circle for a problem involving two standard normal variables. For estimation of failure probability, the impact of sampling points outside the circle generally diminishes as their distance from the center of the sampling area increases. Therefore, a hypersphere with a suitable radius $r_c$ can be pre-established, and a penalty term that escalates with the sample's distance from the center can be introduced to minimize the likelihood of selecting points outside the hypersphere. The penalty coefficient $p$ is set according to the range of LSF $G(\mathbf{u})$ within the sampling space. In this work, the penalty coefficient is set as

$$p = \alpha_s \frac{\bar{G}(\mathbf{u})\big|_{\mathbf{u}\in\mathbf{DoE}} - \underline{G}(\mathbf{u})\big|_{\mathbf{u}\in\mathbf{DoE}}}{4.0} \tag{17}$$

where $\underline{G}(\mathbf{u})\big|_{\mathbf{u}\in\mathbf{DoE}}$ and $\bar{G}(\mathbf{u})\big|_{\mathbf{u}\in\mathbf{DoE}}$ represent the minimum and maximum of the LSF values obtained from the current set of DoE, $\alpha_s = \sqrt{2/N_D}$ refers to the dimensionality adjustment parameter that consider the number of random variables. It should be noted that the setting of penalty coefficient does not stringently confine representative sample points within the specified hypersphere. Provided that the value of the objective function is sufficiently low, samples located outside the hypersphere can also be selected as representative points. This approach circumvents the potential absence of truly representative sample points due to the constraints of the hypersphere.

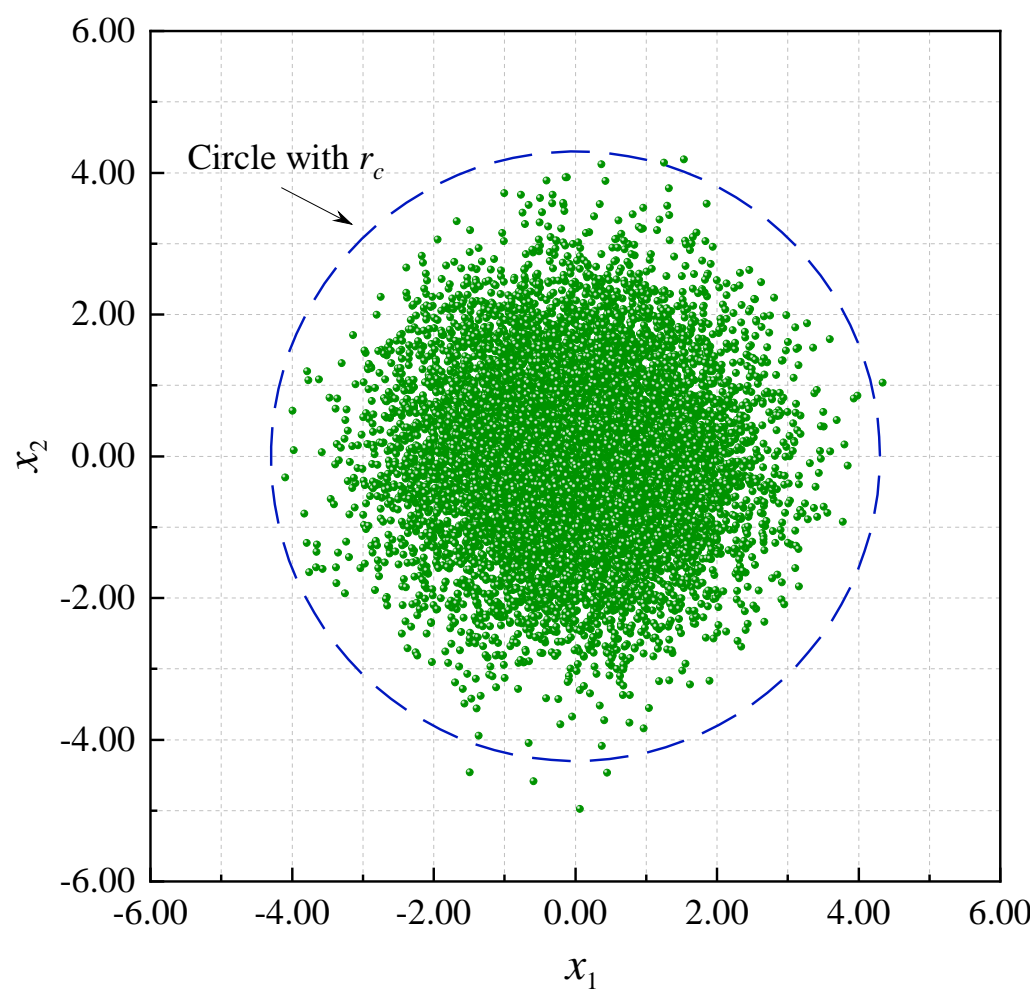

**Fig. 3**. Sampling points of two standard normal variables.

3.2.2. Implementation of the improved AK-MCS

Different from the regular CSP-based AK-MCS where the CSP varies with Monte Carlo population, the improved CSP-free AK-MCS is implemented in two stages: surrogate model construction and MCS with the established surrogate model for the estimation of failure probability. The construction of surrogate model is a two-level iterative process. The outer iteration involves gradual augmentation of the DoE data and the enhancement of the Kriging surrogate model. The inner iteration encompasses the implementation of PSO. The maximum iteration number serves as the stopping condition for the inner iteration, while $\min\left(U\left(\mathbf{u}\right)\right)\geq 2.0$ is the stopping criterion for the outer iteration. To avoid the premature stop of the surrogate model construction, it is required that the DoE set includes at least one sample point in the vicinity of the LSE. The condition to fulfill this requirement is expressed as

$$G\left(\mathbf{u}\right)<0.1\cdot\left\|\nabla\hat{G}\left(\mathbf{u}\right)\right\| \tag{18}$$

where $\nabla\hat{G}\left(\mathbf{u}\right)$ represents the gradient vector of the predicted LSF to the random variables, which can be obtained by using the Kriging model.

The main steps of the improved CSP-free AK-MCS for reliability analysis are listed as follows:

**Step 1**: Set control parameters for PSO, including $N_{swarm}$, $N_{ite_\max}$, $\omega$, $c_1$, $c_2$ and $v_{\max}$.

**Step 2**: Build initial data of DoE. Generate initial set of DoE $\mathbf{S}_{\text{DoE}}=\left(\mathbf{u}^1,\mathbf{u}^2,\ldots,\mathbf{u}^m\right)$ using Latin hypercube sampling, and obtain the LSF values of them denoted as $\mathbf{Y}_{\text{DoE}}=\left(G\left(\mathbf{u}^1\right),G\left(\mathbf{u}^2\right),\ldots,G\left(\mathbf{u}^m\right)\right)$ with $m$ being the number of samples in DoE.

**Step 3**: Construct the Kriging model according to the data in $\mathbf{S}_{\mathrm{DoE}}$ and $\mathbf{Y}_{\mathrm{DoE}}$.

**Step 4**: Based on the current Kriging model, select a representative sample through PSO, which includes the following sub-steps:

Step 4-(1): Initialization of particle swarm. Generate particle swarm by randomly initializing the position and velocity of each particle in the solution space. Evaluate all particles in the swarm using Eqs. (16) and set the current position as the historical best position for each particle. Meanwhile, determine the swarm's best position according to Eq. (15).

Step 4-(2): Update the position and velocity for all particles using Eqs. (12) and (13) and obtain a new generation of particle swarm.

Step 4-(3): Evaluate all particles in the swarm using Eqs. (16), and update the historical best position for each particle and the best position of the whole swarm using Eqs. (14) and (15). If $N_{ite_\max}$ is not reached, turn to Step 4-(2); otherwise, go to the next step.

Step 4-(4): Output the best position of the swarm as the selected representative sample point.

**Step 5**: Receive the representative sampling point $\mathbf{u}^* = \mathbf{u}_{gbest}^{(N_{ite_\max})}$ selected by PSO. If $U(\mathbf{u}^*) \geq 2.0$, go to **Step 6**; otherwise, call the LSF $G(\mathbf{u}^*)$ and add $\mathbf{u}^*$ and $G(\mathbf{u}^*)$ to $\mathbf{S}_{\mathrm{DoE}}$ and $\mathbf{Y}_{\mathrm{DoE}}$, respectively, then turn to **Step 3**.

**Step 6**: Generate Monte Carlo population $\mathbf{S}_{MC} = (\mathbf{u}^1, \mathbf{u}^2, \ldots, \mathbf{u}^{n_{MC}})$ by Monte Carlo sampling.

**Step 7**: Predict the mean values of LSF $u_{\hat{G}}(\mathbf{u}^i)(i = 1, 2, \ldots, n_{MC})$ for all samples in the Monte Carlo population $\mathbf{S}_{MC} = (\mathbf{u}^1, \mathbf{u}^2, \ldots, \mathbf{u}^{n_{MC}})$ using the established Kriging surrogate model. Then, the estimated failure probability $\hat{P}_f$ can be calculated using Eq. (9).

**Step 8**: Check the sufficiency of Monte Carlo population through the variation coefficient of failure probability $V_{\hat{P}_f}$ obtained using Eq. (11). If $V_{\hat{P}_f} < 0.05$, output $\hat{P}_f$ and end the program. If $V_{\hat{P}_f} \geq 0.05$, generate a new batch of samples by Monte Carlo sampling, add these samples to Monte Carlo population $\mathbf{S}_{MC}$, and then turn to **Step 7**.

**Fig. 4** displays the flow chart of the improved CSP-free AK-MCS. It can be observed that the improved AK-MCS eliminates the need for constructing a CSP. Furthermore, there is no requirement for repeated estimation of failure probability during surrogate model construction. This method clearly separates surrogate model construction and failure probability estimation into two distinct stages, thereby facilitating a more transparent implementation process. It is important to emphasize that the proposed method selects representative samples based on an optimization algorithm rather than a CSP-based method. This effectively addresses the challenge of obtaining representative sample points in regions of sparse distribution. With the method proposed, a Kriging surrogate model for reliability analysis with a small failure probability can be constructed without the need to introduce more complex sampling techniques such as subset simulation [9]. This provides a more concise and straightforward approach for reliability analysis.

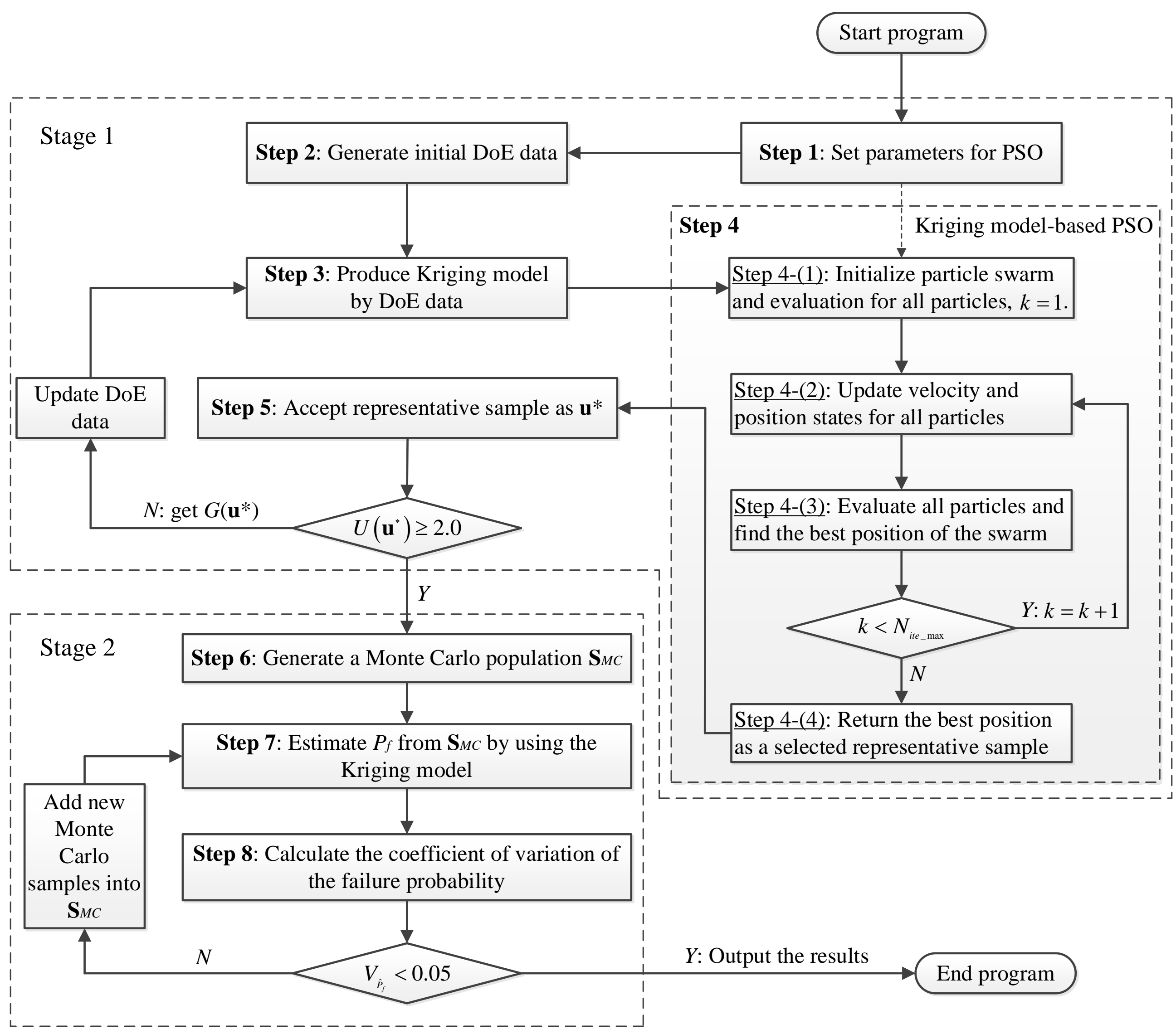

**Fig. 4**. Flow chart of the improved CSP-free AK-MCS.

The representative sample selection and Monte Carlo sampling in the improved AK-MCS are both carried out in standard normal space. The random variables expressed in the original design space can be obtained from those in standard normal space through the Nataf transformation $\mathbf{X}=T(\mathbf{U})$. The transformation of random variables from standard normal space to the original design space is implemented by referring to the open-source package FERUM [49]. Thus, the failure probability is rewritten as [39]

$$P_f=\int I_{G<0}\left(G(\mathbf{u})\right) f_{\mathbf{U}}(\mathbf{u})\mathrm{d}\mathbf{u} \tag{19}$$

where $f_{\mathbf{U}}(\mathbf{u})$ is the joint standard normal distribution density function, $I_{G<0}\left(G(\mathbf{u})\right)$ is the failure indicator determined by the LSE ($G(\mathbf{u})=0$). The relationship between standard normal space and original design space is shown in **Fig. 5**. Based on the standard normal space, the significance of the distance between two sample points is independent on the actual distribution of random variables, which is more appropriate to define the distance parameters for construction of learning functions. In standard normal space, the point on the LSE ($G(\mathbf{u})=0$) with a smaller distance from the origin in standard normal space represents more important information in estimating the true failure probability, such as the MPP shown **Fig. 5**. Based on the reliability theory, the first-order reliability index $\beta$ can be obtained according to the distance from MPP to the coordinate origin as $\beta=\left\|\mathbf{u}_{\mathrm{MPP}}\right\|$.

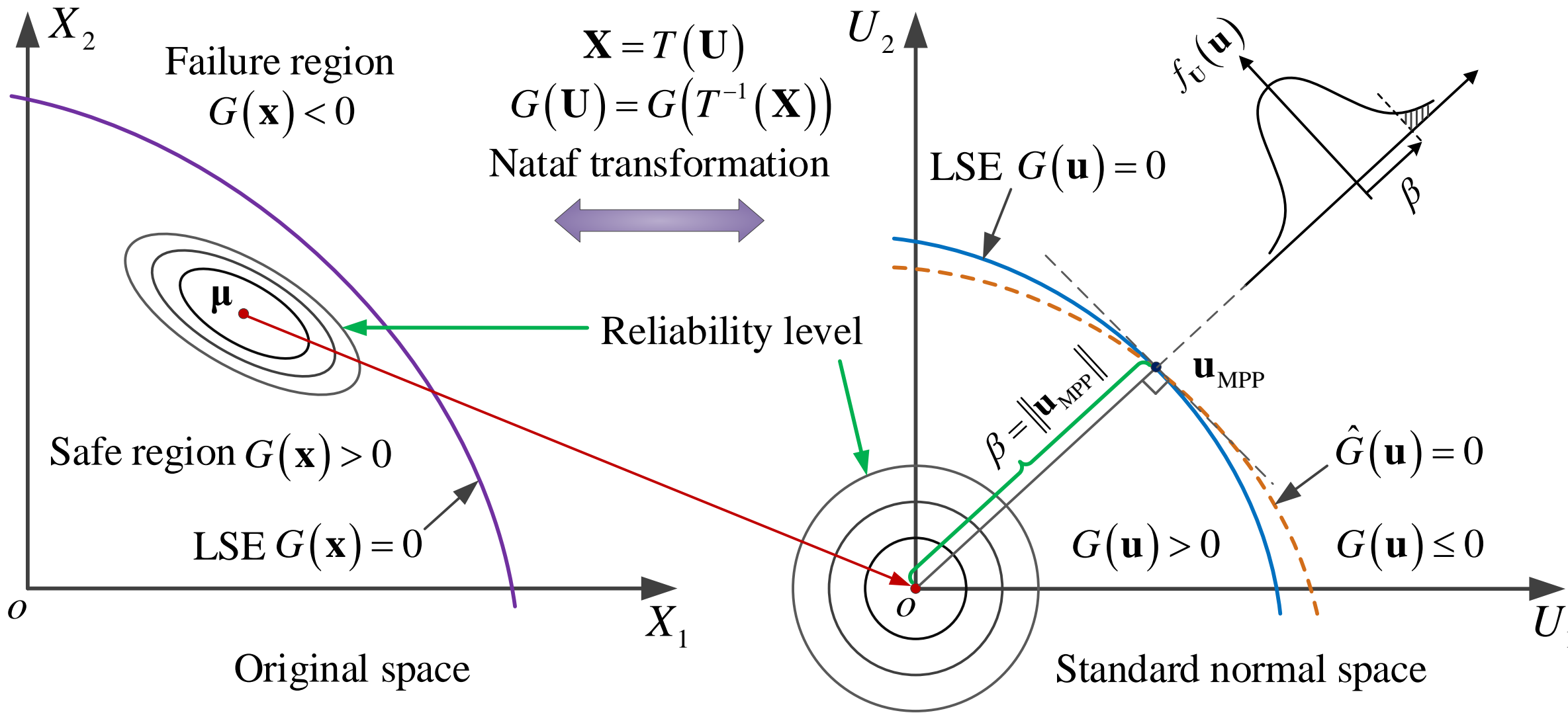

**Fig. 5**. Relationship between original space **X** and standard normal space **U**.

3.2.3. Modification of the objective function

The computational framework of the improved AK-MCS presented in **Sec. 3.2.2** necessitates the resolution of two critical issues. The first issue arises from the neglect of the impact of sample point positions during the evaluation process, which could potentially compromise the efficiency of surrogate model construction. The second issue arises from the representative samples chosen based on the U-learning function, which may display a concentrated distribution in certain areas, thereby expanding the scale of the DoE. In response to the two issues, the following improvements are proposed.

(1) Penalty intensity control

For the sampling points located on the LSE, their importance for reliability analysis and surrogate model construction varies depending on their distance from the sampling center. Typically, points in closer proximity to the sampling center bear greater significance. The objective function delineated in Eq. (16) fails to account for the positional impact of the sample points, despite the incorporation of a penalty term for sample points located outside the hypersphere. Consequently, the additional penalty term is introduced to the objective function, enabling the reformulation of the optimization equation as follows

$$\begin{aligned}
&\text{find} \quad \mathbf{u}^* = \left(u_1^*, u_2^*, \ldots, u_{N_D}^*\right) \\
&\min \quad F_{\text{obj}}\left(\mathbf{u}\right) = \frac{\left\|\mu_{\hat{G}}\left(\mathbf{u}\right)\right| - \delta\right|}{\sigma_{\hat{G}}\left(\mathbf{u}\right)} + p \cdot \max\left(\left\|\mathbf{u}\right\| - r_c, 0\right) + p \cdot \max\left(\left\|\mathbf{u}\right\| - r, 0\right) \\
&\qquad \text{s.t.} \quad u_j \in \left[-u_{\lim}, u_{\lim}\right] \left(j = 1, 2, \ldots, N_D\right)
\end{aligned} \tag{20}$$

where $r$ is a parameter that determines the distribution of penalty intensity. At the beginning of surrogate model construction, $r$ is obtained from the initial DoE set $\mathbf{S}_{\text{DoE_init}}$ as

$$r = r_0 = \frac{1}{m}\sum_{i=1}^{m}\left\|\mathbf{u}^i\right\| \quad \left(\mathbf{u}^i \in \mathbf{S}_{\text{DoE_init}}\right) \tag{21}$$

During the construction of surrogate model, $r$ is determined as

$$r = \min\left(\max\left(\left\|\mathbf{u}^i\right\|\right), r_c\right) \quad \left(\mathbf{u}^i \in \mathbf{S}_{\text{DoE}}, \mathbf{u}^i \notin \mathbf{S}_{\text{DoE_init}}\right) \tag{22}$$

Then, the stopping criterion of surrogate model construction is modified as $\min\left(U\left(\mathbf{u}\right)\right) \geq 2.0$ and $r = r_c$. For clarity, the sketch map of penalty intensity distribution in the objective function with two design variables is shown in **Fig. 6**.

As mentioned above, $r$ increases gradually during the construction of the surrogate model. From the perspective of calculation logic, under the premise that there are enough sample points near the sampling center, it is necessary to gradually relax the restrictions on sample points far from the sampling center in order to further capture points on LSE and establish an optimization process that expands the search range. This improvement adaptively expands the range of obtaining representative sample points during surrogate model construction, increasing the likelihood of obtaining representative sample points and ensuring the stability of DoE. The effectiveness of this improvement will be verified in the validation section.

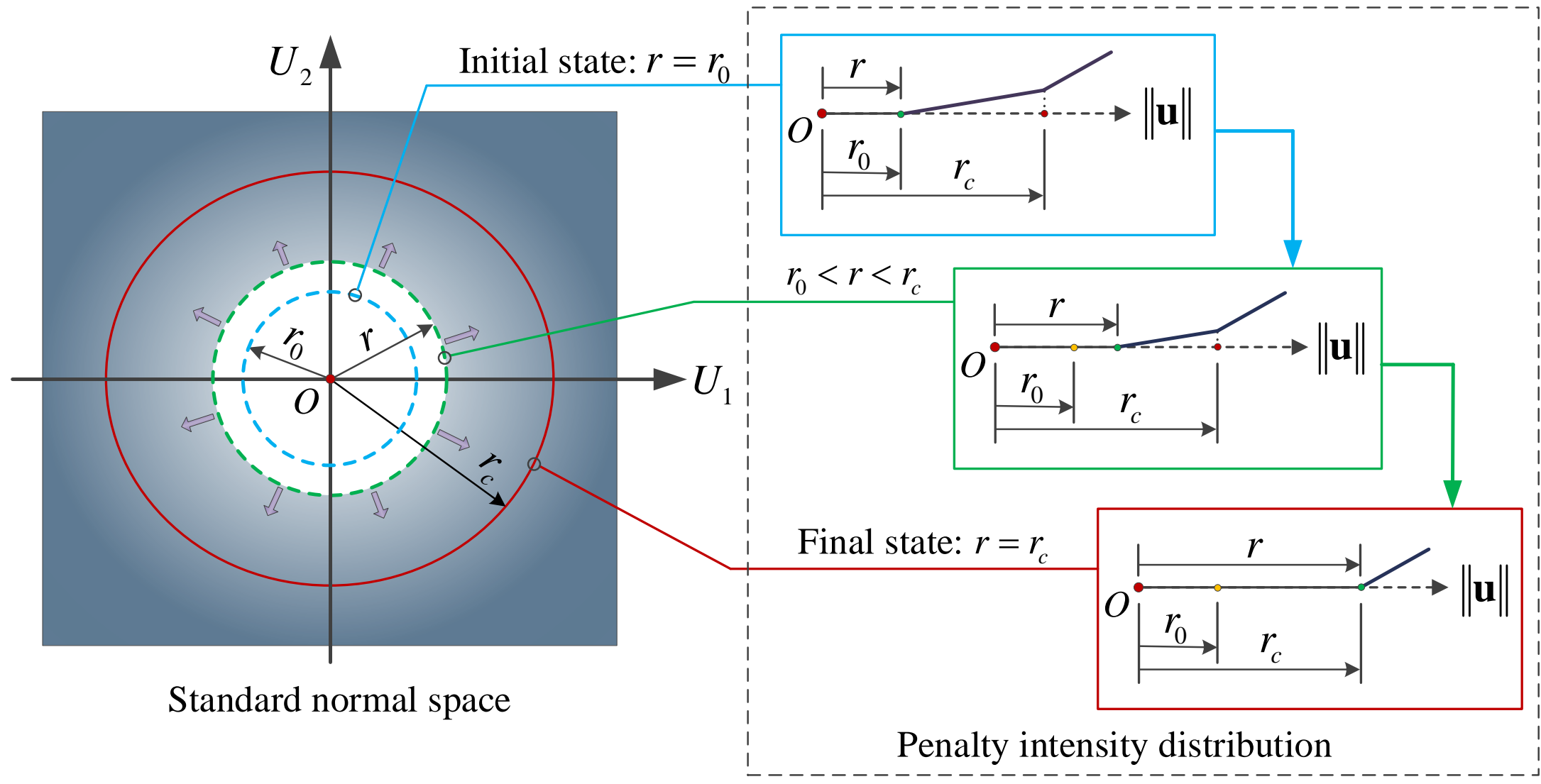

**Fig. 6**. Penalty intensity distribution in the improved AK-MCS.

(2) Density control

The optimization capability of PSO allows the improved AK-MCS to generate new sample points with lower value of $U(\mathbf{u})$ compared to the CSP-based AK-MCS. Consequently, it poses a greater challenge for the representative samples generated to meet the stopping criterion of $\min(U(\mathbf{u})) \geq 2.0$, potentially increasing the size of DoE. Furthermore, according to the definition of the U-learning function, there may be points within the designated design space with an extremely low value of $U(\mathbf{u})$, despite their minimal contribution towards enhancing the accuracy of the surrogate model. In the improved AK-MCS, these points may be identified through optimization algorithms, thereby increasing the size of DoE.

This issue is elucidated using the LSF $G(x) = \sin(x), x \in [\pi/2, 5\pi/2]$ for enhanced comprehension. Assuming that the determined samples of DoE are located at $x$ = 4.0, 6.18, 6.38, and 7.5, a Kriging model is constructed based on these samples, and the subsequent representative sample point is located according to the current Kriging model. **Fig. 7** displays the curves of $\mu_{\hat{G}}(x)$, $\mu_{\hat{G}}(x) \pm \sigma_{\hat{G}}(x)$ and $U(x)$ (expressed by $\log_{10} U(x)$) based on the current Kriging model. It is observable that the minimum value of $U(x)$ appears proximate to $x^* = 2\pi$ in the current state. In other words, despite the DoE samples already encompassing the two points near $x^* = 2\pi$ ($x = 6.18$ and $x = 6.38$), resulting in a significantly small Kriging variance $\sigma_{\hat{G}}^2(x)$ within this narrow range, the minimum value remains within this range (point $A$ in the figure) due to the existence of $\mu_{\hat{G}}(x) = 0$. An effective optimization algorithm can identify point $A$ and add it to DoE. However, when points at $x = 6.18$ and $x = 6.38$ are included in DoE, the contribution of point $A$ to the Kriging model is quite limited. In fact, for the current Kriging model shown in **Fig. 7**, point $B$ is the truly valuable sample point due to its high Kriging variance and low absolute value of the predicted mean $|\mu_{\hat{G}}(x)|$. Presuming that point $B$ can be selected and added to DoE, the points in proximity to point $C$ can be subsequently obtained based on the learning function and then the established Kriging model performs well in distinguishing between failure and safe regions.

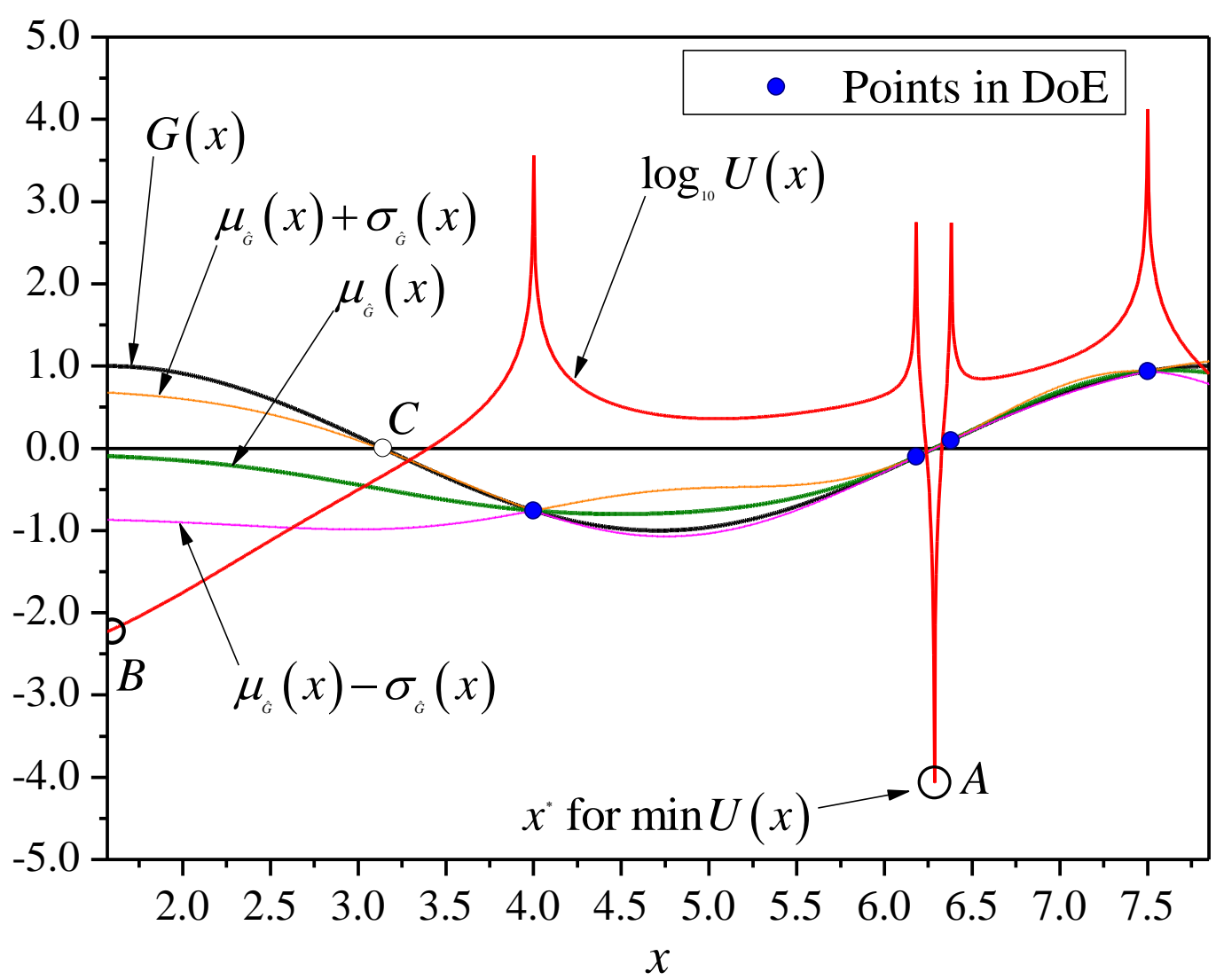

**Fig. 7**. Discussion on the learning function $U$.

To circumvent the aforementioned problem, it is suggested that any subsequent representative sample identified should preserve a specific distance from the existing DoE sample points. Consequently, a density control mechanism for distance constraint is implemented to prevent an excessively dense congregation of sample points within the DoE. This allows the optimization formulation to be restructured as follows

$$\begin{aligned}
&\text{find} \quad \mathbf{u}^* = \left(u_1^*, u_2^*, \ldots, u_{N_D}^*\right) \\
&\min \quad F_{\text{obj}}\left(\mathbf{u}\right) = \frac{\left\| \mu_{\hat{G}}\left(\mathbf{u}\right) \right| - \delta \right|}{\sigma_{\hat{G}}\left(\mathbf{u}\right)} + p \cdot \max\left(\left\|\mathbf{u}\right\| - r_c, 0\right) + p \cdot \max\left(\left\|\mathbf{u}\right\| - r, 0\right) + 2.0c \\
&\qquad \text{s.t.} \quad u_j \in \left[-u_{\lim}, u_{\lim}\right]\left(j = 1, 2, \ldots, N_D\right)
\end{aligned} \tag{23}$$

where

$$c = \begin{cases} 0, & \alpha_s \left\|\mathbf{u} - \mathbf{u}^i\right\| \geq d_c \quad \text{for } \mathbf{u}^i \in \mathbf{S}_{\text{DoE}} \\ 1.0, & \text{otherwise} \end{cases} \tag{24}$$

In Eq. (24), $d_c$ represents the density control index, and it is determined as

$$d_c = \begin{cases} r_d, & \alpha_s \left\|\mathbf{u}\right\| \leq \lambda r_d^{-1} \\ \dfrac{\lambda}{\alpha_s}\left\|\mathbf{u}\right\|^{-1}, & \alpha_s \left\|\mathbf{u}\right\| > \lambda r_d^{-1} \end{cases} \tag{25}$$

where $r_d$ and $\lambda$ represent the two parameters that characterize the distance constraint. The modified objective function in Eq. (23) can be interpreted as augmenting the learning function value by 2.0 when the distance between the position of point $\mathbf{u}$ and any sample $\mathbf{u}^i$ in the current DoE is less than the density control index $d_c$. For elucidation, the delineation of the density control index is depicted in **Fig. 8**. According to the definition in Eq. (25), the value of density control index $d_c$ diminishes in correlation with the amplification of the distance between the sample point and the sampling center. The rationale behind this configuration is to alleviate the density restriction of DoE samples in regions distanced from the sampling center, thereby enabling the acquisition of denser DoE sample points in these areas. This ensures that the surrogate model has adequate precision for problems with small failure probabilities.

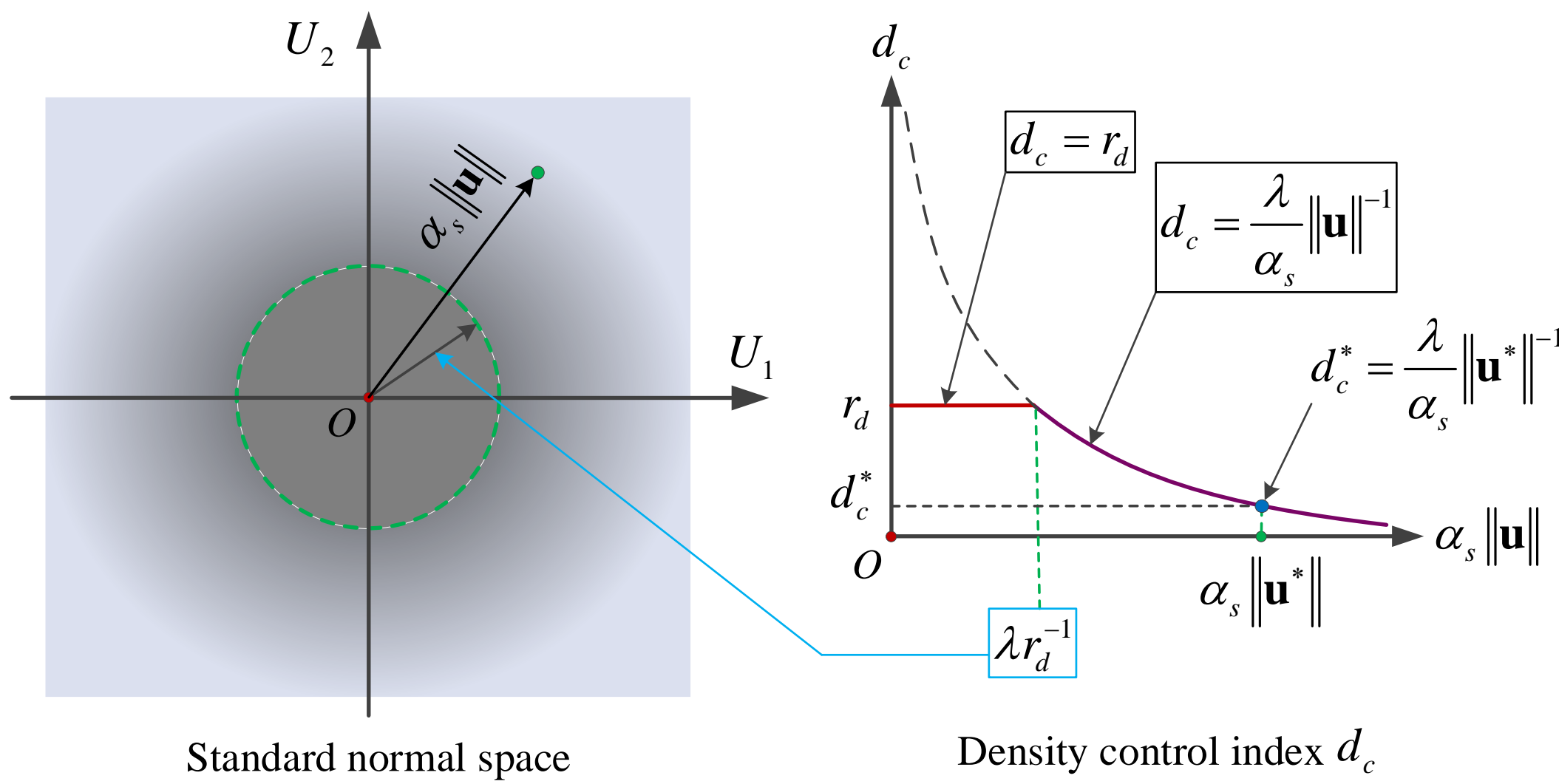

**Fig. 8**. Density control index for DoE.

3.2.4. Configuration of the control parameters

The control parameters encompassed in the objective functions include the small quantity $\delta$, the radius of the hypersphere $r_c$ and the two parameters delineating the distance constraint. The principles of their configuration are as follows.

Typically, the small quantity $\delta$ is set to 0.001 to maximize the acquisition of representative samples close to LSE. In certain exceptional scenarios, to prevent difficulties in terminating the construction of surrogate model due to repeated acquisition of points that are extremely close to the LSE, and to predict the gradient values of the LSF in the vicinity of limit state more accurately, $\delta$ can be assigned a small positive value within the range of (0.001,0.2].

The determination of $r_c$ should be based on the number of random variables and the possible range of failure probability. Usually, the relationship between the proportion of samples outside the hypersphere and the value of $r_c$ can be pre-determined for a reliability problem with $n$ random variables, by generating a certain scale of samples in the standard normal space. The suitable setting of $r_c$ can be determined based on the principle of matching the proportion mentioned above with the potential value of failure probability. To enhance its operationality, $r_c$ can be set to $r_c = \sqrt{n} + \gamma$, where $\gamma$ is a positive value with the range of [0, 4]. Note that the proportion of samples outside the hypersphere with $\gamma = 0$ is close to 50% for problems with a sufficient number of random variables. Under the setting of $\gamma = 3.0$, the order of magnitude for the proportion can be reduced to $1.0 \times 10^{-5}$.

In the context of the distance constraint, the parameters $r_d$ and $\lambda$ assume distinct roles. $r_d$

is utilized to establish the condition of the distance constraint for the sampling center area, while $\lambda$ is usually set to determine the distance constraint in the sampling edge area. Generally, the setting of $r_d = \lambda = 0.5$ can applied to most reliability problems. To strengthen the distance constraint, they can be configured with larger values. Exceptionally, for the LSF that is complex in the sampling center area, the value of $r_d$ should be decreased to describe the failure regions with greater accuracy.

## 4. Academic validation

This section conducts academic validation of the proposed method. Initially, the efficacy of the algorithm employed in the proposed method is investigated through a classical series system with four branches and its variations. Subsequently, the computational performance of the proposed method is further evaluated through a series of benchmark problems.

For ease of reference, the improved CSP-Free AK-MCS based on Eq. (16), Eq. (20) and Eq. (23) are denoted as CFAK-MCS$^b$, CFAK-MCS$^p$ and CFAK-MCS$^c$, respectively. The superscripts b, p and c on CFAK-MCS correspond to the basic form, the form with penalty intensity control only, and the complete form with both penalty intensity control and density control.

For PSO used in the CFAK-MCSs, the control parameters $N_{ite_max}$, $v_{max}$, $\omega$, $c_1$ and $c_2$ are set to $N_{ite_max} = 50$, $v_{max} = 0.3$, $\omega = 0.729$ and $c_1 = c_2 = 2.0$ throughout the study. In the investigation for effectiveness of PSO (**Sec 4.1.1**), the size of particle swarm is set to $N_{swarm} = 50$ for the purpose of better illustrating the alterations in particle position, while $N_{swarm} = 100$ for other investigations and the study of benchmark problems. In **Sec. 4.1**, the control parameters used in the objective function are set to $r_c = 4.3$, $r_d = \lambda = 0.5$ and $\delta = 0.001$, and the number of initial DoE samples for surrogate model methods is set to 6. For the study of benchmark problems, the settings of control parameters for the objective function and the number of initial DoE samples are provided at the beginning of **Sec. 4.2**.

To guarantee the dependability of the outcomes, it is necessary to run the computation program multiple times under the specified control parameter settings. Subsequently, the computational performance of a reliability method is evaluated via some of the indicators introduced as follows.

(1) The average estimated failure probability $\hat{P}_f$ for the *n* runs, which is obtained by taking the mean of the estimated failure probability from each individual run. This is expressed mathematically as

$$\hat{P}_f = \frac{1}{n}\sum_{i=1}^{n}\hat{P}_f^{(i)} \tag{26}$$

where $\hat{P}_f^{(i)}$ represents the estimated failure probability from $i^{th}$ run. As for a single run, the average estimated failure probability is equivalent to the estimated failure probability for that specific run.

(2) The relative error $\varepsilon_{\hat{P}_f}$ of the estimated failure probability, which is defined as

$$\varepsilon_{\hat{P}_f} = \frac{\left|\hat{P}_f - P_{f,ref}\right|}{P_{f,ref}} \times 100\% \tag{27}$$

where $P_{f,ref}$ is the reference failure probability obtained by MCS or an analytical method, and $\hat{P}_f$ represents the (average) estimated failure probability obtained by the reliability method to be evaluated.

(3) The average relative error $\bar{\varepsilon}_{\hat{P}_f}$ for the *n* runs, which is used for the situation where the reference failure probability is repeatedly obtained in each run. It is defined as

$$\bar{\varepsilon}_{\hat{P}_f} = \frac{1}{n}\sum_{i=1}^{n}\left(\frac{\left|\hat{P}_f^{(i)} - P_{f,ref}^{(i)}\right|}{P_{f,ref}^{(i)}}\right) \times 100\% \tag{28}$$

where $P_{f,ref}^{(i)}$ refers to the reference failure probability obtained by MCS for the $i^{\text{th}}$ run.

(4) The average number of calls to LSF for the *n* runs, which is denoted as $\bar{N}_{\mathrm{G}}$ and calculated by averaging the number of calls to LSF, $N_{\mathrm{G}}$, for each run.

(5) The standard deviation of the failure probabilities for the *n* runs, which is denoted as $\sigma_{\hat{P}_f}$.

(6) The average CPU time for the *n* runs, which is denoted as $\bar{T}$.

(7) The average number of LSF predictions using the surrogate model during surrogate model construction for the *n* runs, which is denoted as $\bar{N}_{\mathrm{pred}}$.

Generally, the average estimated failure probability is considered as the primary outcome of a reliability method. The relative error or average relative error serves as the indicators to evaluate the solution accuracy of a reliability method. The average number of calls to LSF and the average CPU time are employed to evaluate the efficiency of a reliability method.

The proposed method has been programmed in MATLAB and run on a computer having an Intel® Core™ i7-8700 processor and a CPU at 3.2GHz with 64GB of RAM.

*4.1. Investigation of algorithm effectiveness*

This section investigates the effectiveness of the algorithm employed in the proposed method through a classical series system with four branches [16] and its variations. The LSF of the system with four branches is defined in Eq. (29), where $x_1$ and $x_2$ are independent standard normal distributed random variables.

$$G(x_1,x_2)=\min\begin{Bmatrix}3.0+0.1\times(x_1-x_2)^2-(x_1+x_2)/\sqrt{2}\\3.0+0.1\times(x_1-x_2)^2+(x_1+x_2)/\sqrt{2}\\(x_1-x_2)+6.0/\sqrt{2}\\(x_2-x_1)+6.0/\sqrt{2}\end{Bmatrix} \tag{29}$$

In this section, the investigation includes three aspects: (1) The efficacy of PSO in the construction of the active learning Kriging surrogate model, (2) The influence of penalty intensity control and density control, and (3) The performance of the algorithm in cases with small failure probabilities.

4.1.1. Effectiveness of PSO

The failure probability of the system is assessed using both MCS and CFAK-MCS$^b$. In MCS, the size of Monte Carlo population is set to $1\times10^6$ and the population is generated via random sampling. This Monte Carlo population is also employed for estimating the failure probability in the second stage of CFAK-MCS$^b$. **Table 1** presents the results derived from both MCS and CFAK-MCS$^b$. As indicated by **Table 1**, CFAK-MCS$^b$ attains a satisfactory level of accuracy. The number of calls to LSF in CFAK-MCS$^b$ is 171, and the relative error of the failure probability is a mere 0.057%.

**Table 1** Failure probability of the series system with four branches (single run).

| Method | $\hat{P}_f$ /10$^{-2}$ | $\varepsilon_{\hat{P}_f}$ /% | $N_{\mathrm{G}}$ |
|---|---|---|---|
| MCS | 0.4341 | - | $1\times10^6$ |
| CFAK-MCS$^b$ | 0.4339 | 0.057 | 171 |

**Fig. 9** illustrates the changes of the predicted LSE determined by CFAK-MCS$^b$ when the number of samples in DoE, denoted as $N_{\mathrm{DoE}}$, is 21, 36, 51, and 171, respectively. It is observable that, as $N_{\mathrm{DoE}}$ increases, the predicted LSE progressively approximates the true LSE. Ultimately, the failure and safe points are accurately differentiated by the predicted LSE, as shown in **Fig. 10**.

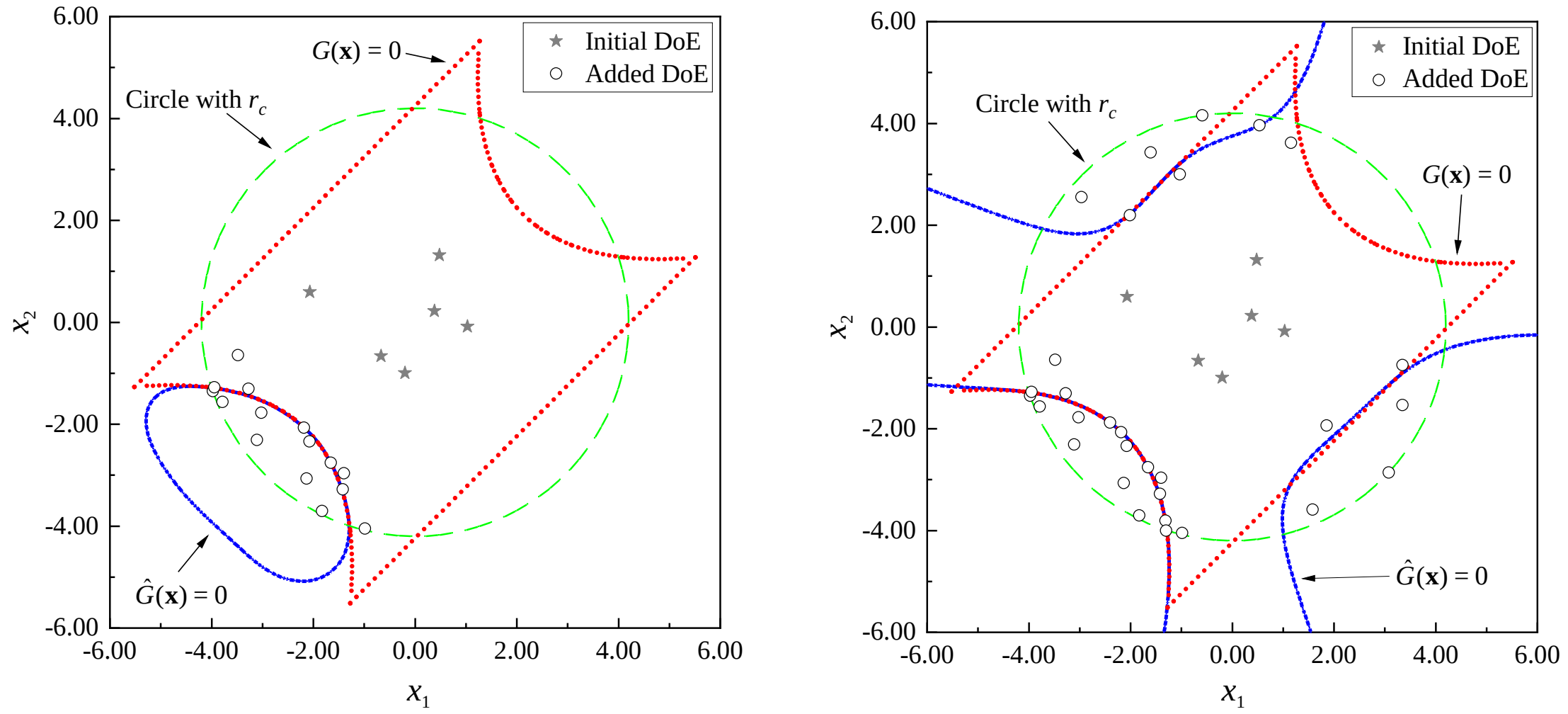

(a) $N_{\mathrm{G}} = 21$ (b) $N_{\mathrm{G}} = 36$

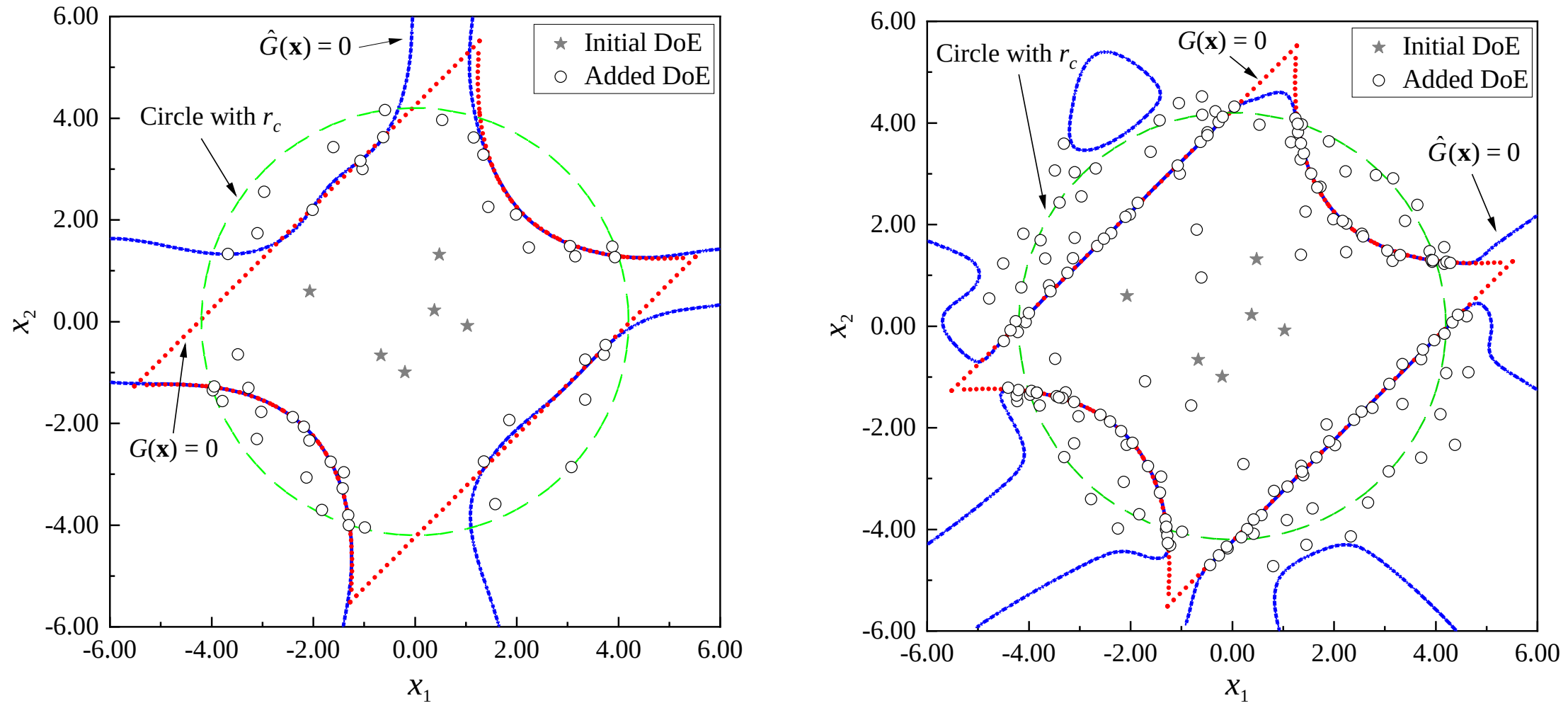

(c) $N_{\mathrm{G}} = 51$ (d) $N_{\mathrm{G}} = 171$

**Fig. 9**. The predicted LSE of CFAK-MCS[b] in evolution process.

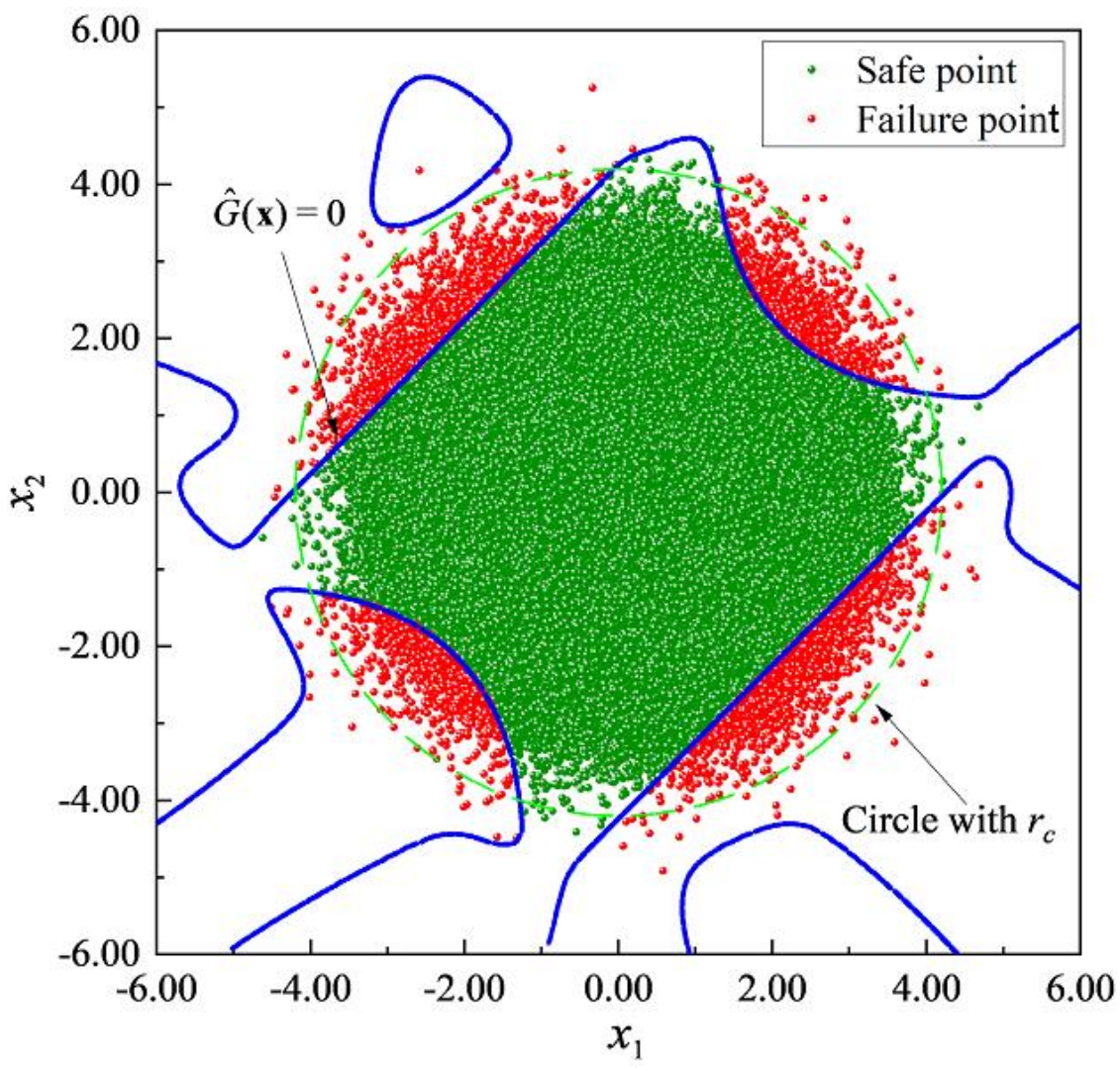

**Fig. 10**. Predicted LSE and distribution of safe and failure points.

In contrast to the CSP-based AK-MCS, CFAK-MCS[b] procures representative samples through an intelligent search based on PSO. **Fig. 11** illustrates the correlation between the sample points in DoE and the objective function value during the construction of the surrogate model in CFAK-MCS[b]. It specifically depicts the distribution of the objective function value when the size of DoE is 21, 36, 51, and 171. The representative sample points (depicted as green points) found using PSO are located in positions where the objective function value is lower (dark blue region) under the prediction of the current surrogate model. This substantiates the effectiveness of PSO optimization process. Furthermore, the selection of a single representative sample point using PSO is explored. **Fig. 12** exhibits the details of obtaining the 21[th] DoE sample point, including the distribution of initially generated particles and their distribution after 15, 30, and 50 iterations. The red arrows depict the motion direction of particles in the subsequent iteration. As iterations progress, particles move closer together and eventually obtain the best position of the swarm in a region with a relatively dense distribution of particles. This signifies the effectiveness of PSO in the proposed method.

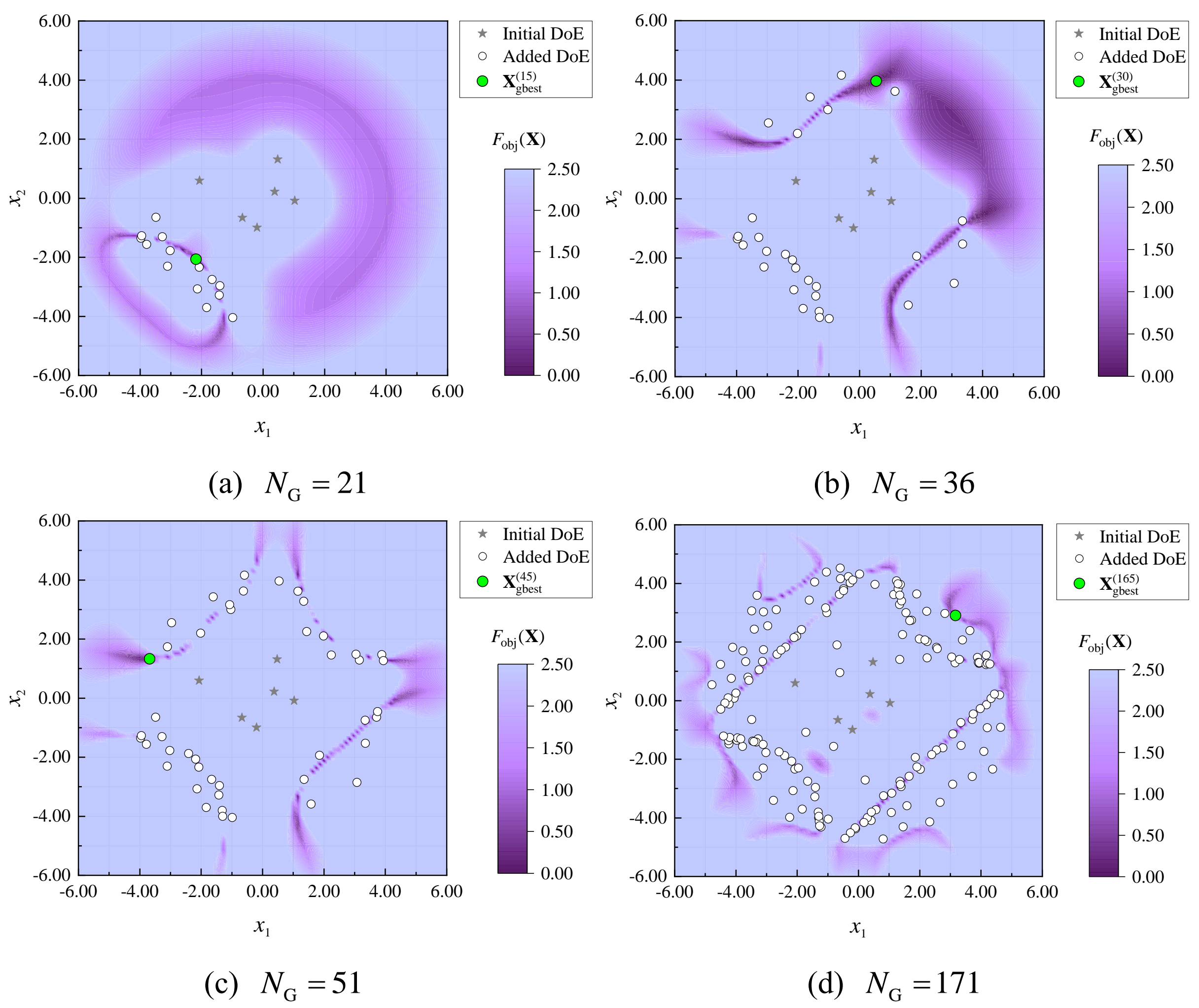

(a) $N_G = 21$ (b) $N_G = 36$

(c) $N_G = 51$ (d) $N_G = 171$

**Fig. 11**. Distribution of objective function and DoE points.

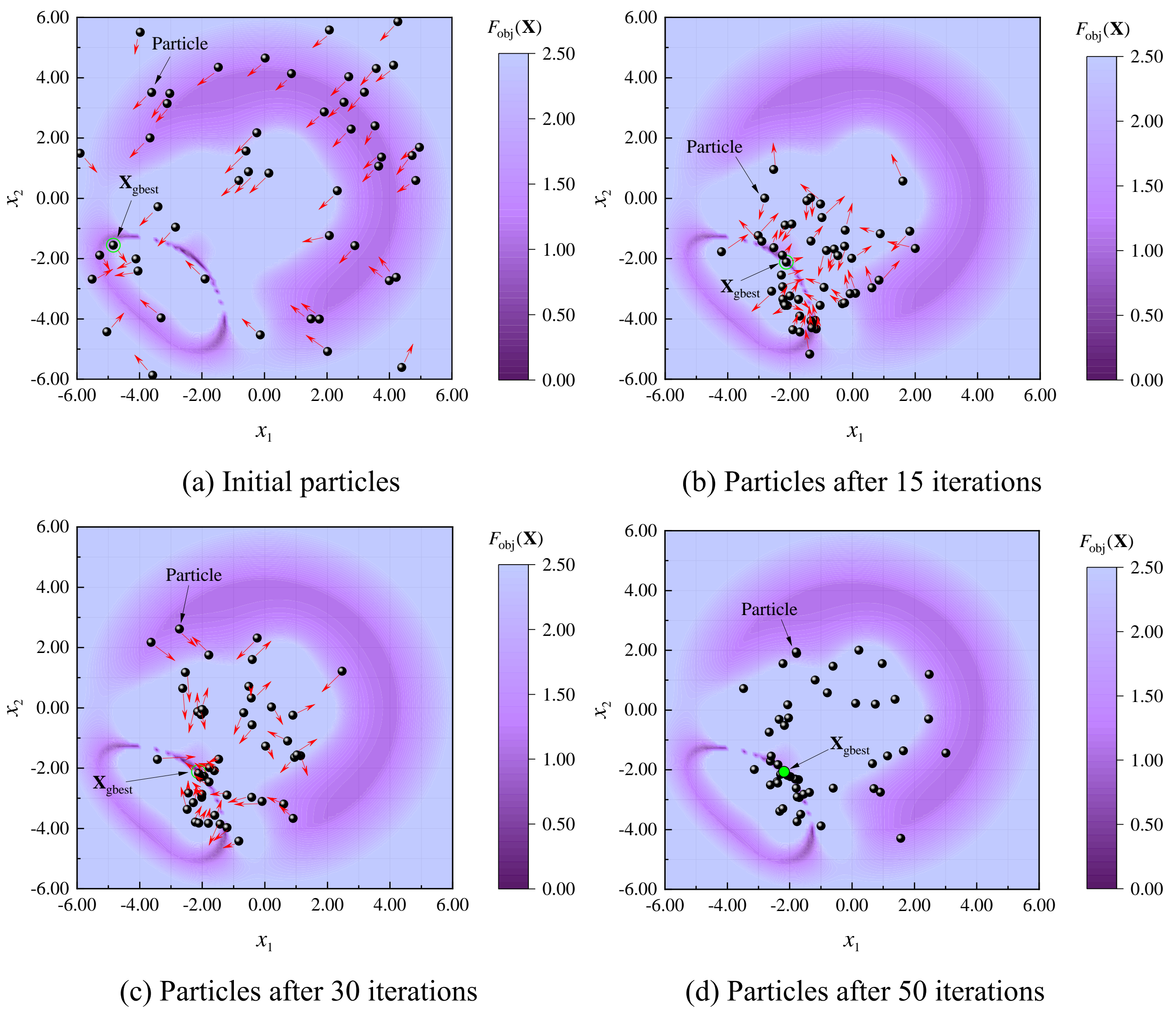

(a) Initial particles

(b) Particles after 15 iterations

(c) Particles after 30 iterations

(d) Particles after 50 iterations

**Fig. 12**. Demonstration of PSO process for the 21th point.

Generally, the optimization capability of PSO increases with the increase of $N_{swarm}$ and $N_{ite_max}$. Using the series system with four branches, an investigation is conducted to ascertain a more reasonable configuration of these two parameters for the subsequent work. To ensure the reference of the results, 30 runs are executed for each parameter setting in this investigation. Initially, **Table 2** presents the average relative error and number of calls to LSF with different setting of $N_{swarm}$ under $N_{ite_max}=50$. The outcomes indicate that as the size of swarm amplifies, both the solution accuracy and the number of calls to LSF increase. Furthermore, under the setting of $N_{swarm}=50$ and $N_{swarm}=100$, **Table 3** and **Table 4** exhibit the results for different setting of $N_{ite_max}$, which reveal that the solution accuracy and the number of calls to LSF both increase with the increase of the maximum iteration number. Based on the results listed in **Table 2**-**Table 4**, considering the ideal equilibrium between solution accuracy and efficiency, the following setting are selected for the subsequent research: $N_{swarm}=100$ and $N_{ite_\max}=50$.

**Table 2** Investigation of PSO parameters under $N_{ite_\max}=50$ (30 runs).

| $N_{swarm}$ | $N_{ite_\max}$ | $\bar{\varepsilon}_{\hat{P}_f}$ /% | $\bar{N}_{\mathrm{G}}$ |
|---|---|---|---|
| 25 | 50 | 0.129 | 159.65 |
| 50 | 50 | 0.047 | 184.50 |
| 75 | 50 | 0.035 | 187.25 |
| 100 | 50 | 0.027 | 198.62 |
| 125 | 50 | 0.026 | 212.95 |

**Table 3** Investigation of PSO parameters under $N_{swarm}=50$ (30 runs).

| $N_{swarm}$ | $N_{ite_\max}$ | $\bar{\varepsilon}_{\hat{P}_f}$ /% | $\bar{N}_{\mathrm{G}}$ |
|---|---|---|---|
| 50 | 25 | 0.100 | 171.80 |
| 50 | 50 | 0.047 | 184.50 |
| 50 | 75 | 0.038 | 192.50 |
| 50 | 100 | 0.036 | 202.65 |

**Table 4** Investigation of PSO parameters under $N_{swarm}=100$ (30 runs).

| $N_{swarm}$ | $N_{ite_\max}$ | $\bar{\varepsilon}_{\hat{P}_f}$ /% | $\bar{N}_{\mathrm{G}}$ |
|---|---|---|---|
| 100 | 25 | 0.062 | 193.95 |
| 100 | 50 | 0.027 | 198.62 |
| 100 | 75 | 0.016 | 204.84 |
| 100 | 100 | 0.015 | 206.47 |

4.1.2. Impact of penalty intensity control and density control

The impact of penalty intensity control and density control on the performance of the CFAK-MCSs is further examined. A collection of results obtained using MCS, CFAK-MCS$^{b}$, CFAK-MCS$^{p}$ and CFAK-MCS$^{c}$ is shown in **Table 5** (single run). The results demonstrate that the number of calls to LSF in CFAK-MCS$^{p}$ and CFAK-MCS$^{c}$ is significantly reduced comparing to that of CFAK-MCS$^{b}$, and the three surrogate model methods exhibit similar solution accuracy. This indicates that the introduction of penalty intensity control can enhance the ability to obtain representative sample points, thereby significantly reducing the size of DoE. Compared to CFAK-MCS$^{p}$, the size of DoE for CFAK-MCS$^{c}$ is further reduced, indicating that density control can further improve computational efficiency. In this set of results, 82 DoE samples are sufficient for CFAK-MCS$^{c}$. To provide a more intuitive understanding, **Fig. 13** displays the DoE and predicted LSE during the construction of surrogate model. As indicated in **Fig. 13**, most of the DoE points are close to the true LSE, so they make a significant contribution to the accuracy of the surrogate model. The predicted LSE obtained from the completed surrogate model is almost consistent with the true one.

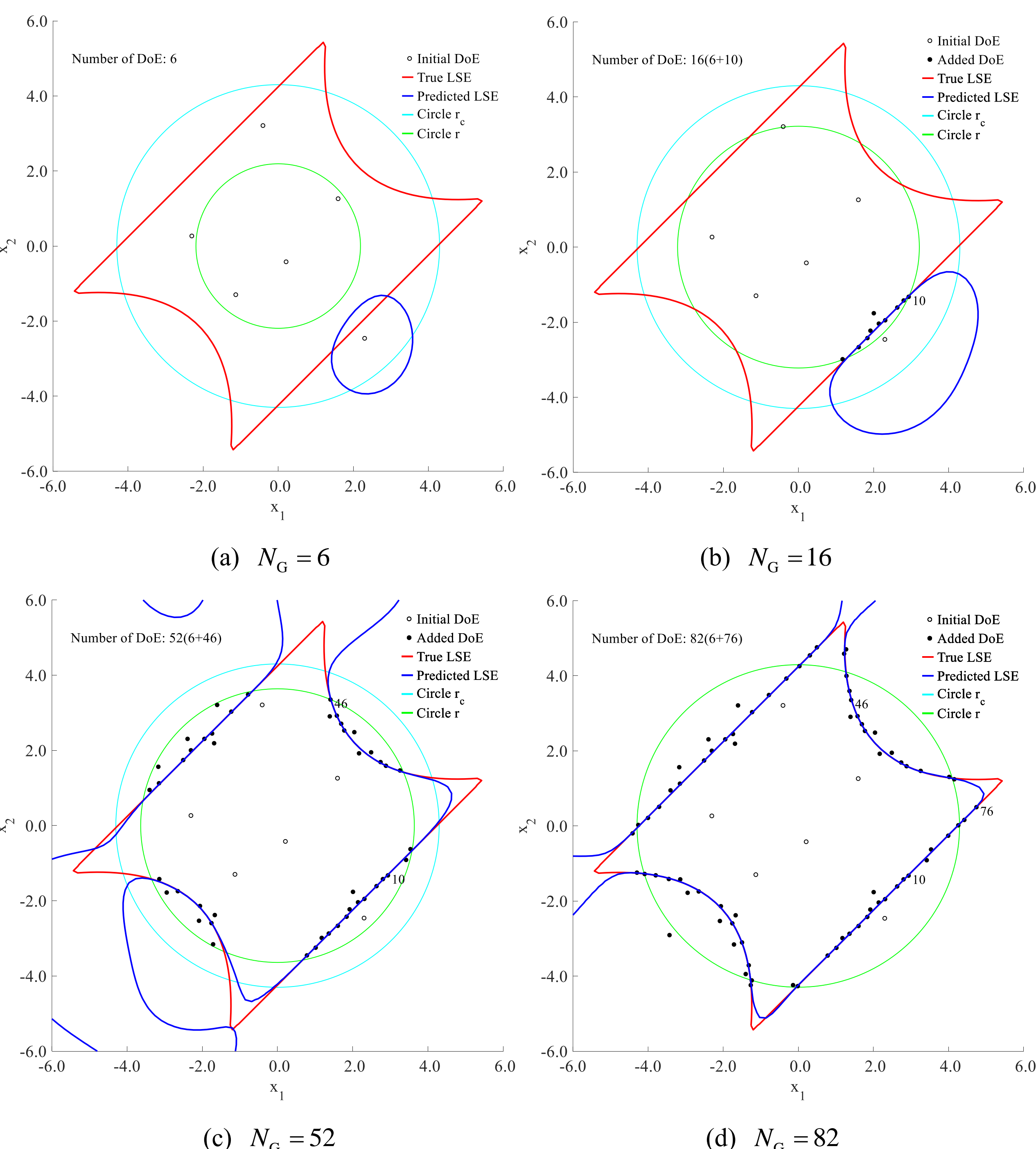

(a) $N_G = 6$ (b) $N_G = 16$

(c) $N_G = 52$ (d) $N_G = 82$

**Fig. 13**. Predicted LSE and the DoE of CFAK-MCS[c].

**Table 5** Results of the series system with four branches obtained by CFAK-MCSs (single run).

| Method | $\hat{P}_f$ /10$^{-3}$ | $\varepsilon_{\hat{P}_f}$ /% | $N_{\mathrm{G}}$ |
|---|---|---|---|
| MCS | 4.4364 | - | $1\times10^{6}$ |
| CFAK-MCS$^{b}$ | 4.4346 | 0.041 | 189 |
| CFAK-MCS$^{p}$ | 4.4334 | 0.068 | 100 |
| CFAK-MCS$^{c}$ | 4.4314 | 0.113 | 82 |

Taking into account the inherent randomness in the proposed method, 50 runs are executed to ensure the reference of the results. **Table 6** lists the statistical results of these 50 runs and the results from the conventional AK-MCS. For a more intuitive understanding, **Fig. 14** illustrates the size of DoE and the relative error of the failure probability for each of the 50 runs conducted using CFAK-MCS$^{c}$. It can be observed that the maximum relative error does not surpass 1.0%, while the maximum size of DoE does not exceed 160.

As indicated by **Table 6**, CFAK-MCS$^{b}$ exhibits extremely high accuracy but a relatively large DOE size. For the three methods of CFAK-MCS$^{b}$, CFAK-MCS$^{p}$, and CFAK-MCS$^{c}$, their computational efficiency improves sequentially, which is reflected in the changes in the size of DOE. The introduction of penalty intensity control and density control can be considered to play a significant role in improving the efficiency of surrogate model construction. Although the solution accuracy may be reduced after the introduction of penalty intensity and density control, the relative errors of failure probability obtained from CFAK-MCS$^{p}$ and CFAK-MCS$^{c}$ are both within a satisfactory range. Compared with the existing AK-MCS methods, CFAK-MCS$^{p}$ and CFAK-MCS$^{c}$ demonstrate higher accuracy. From the perspective of algorithm performance, CFAK-MCS$^{c}$ can achieve a satisfactory balance between solution accuracy and efficiency.

**Table 6** Summary of results for the series system with four branches (50 runs).

| Method | $\hat{P}_f$ /10$^{-3}$ | $\sigma_{\hat{P}_f}$ /10$^{-5}$ | $\overline{\varepsilon}_{\hat{P}_f}$ /% | $\overline{N}_{\mathrm{G}}$ |
|---|---|---|---|---|
| MCS | 4.4454 | 6.9786 | - | $1\times10^{6}$ |
| CFAK-MCS$^{b}$ | 4.4450 | 7.0258 | 0.056 | 182.4 |
| CFAK-MCS$^{p}$ | 4.4436 | 7.1164 | 0.092 | 114.4 |
| CFAK-MCS$^{c}$ | 4.4441 | 6.8252 | 0.131 | 94.8 |
| AK-MCS+U [16] | 4.416 | - | 0.650 | 126 |
| AK-MCS+EFF [16] | 4.412 | - | 0.740 | 124 |

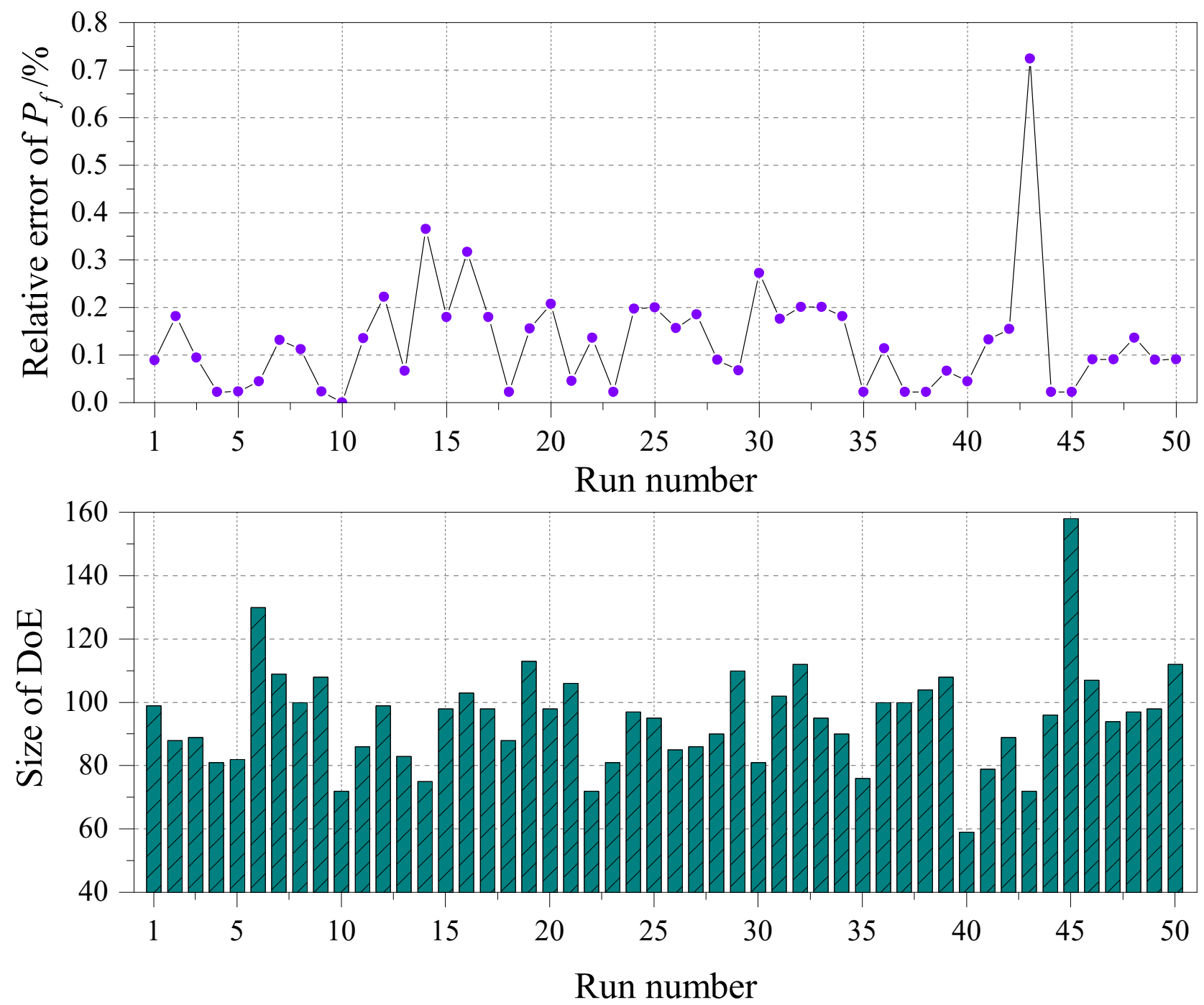

**Fig. 14**. Size of DoE and relative error for CFAK-MCS$^c$ (50 runs).

4.1.2. Small failure probability

The LSF of the system is modified to the following form to investigate the performance of the proposed method under small failure probability.

$$G(x_1, x_2) = \min \begin{Bmatrix} k + 0.1\times(x_1 - x_2)^2 - (x_1 + x_2)/\sqrt{2} \\ k + 0.1\times(x_1 - x_2)^2 + (x_1 + x_2)/\sqrt{2} \end{Bmatrix} \tag{30}$$

where $x_1$ and $x_2$ are independent standard normal distributed random variables, and $k$ is a parameter related to the failure probability. This study evaluates the cases of $k = 3.4$ and $k = 3.9$, the failure probability of the system is obtained using MCS, AK-MCS+U and the CFAK-MCSs.

The Monte Carlo populations are set to $1\times10^6$ and $8\times10^6$ for the cases of $k = 3.4$ and $k = 3.9$, respectively. The Monte Carlo populations for the cases of $k = 3.4$ and $k = 3.9$ are both generated through random sampling, and they are used for estimating the failure probability in various methods. In surrogate model construction of AK-MCS, the size of Monte Carlo population $\mathbf{S}_{MC}$ at the start is set to $\mathrm{size}(\mathbf{S}_{MC}) = 5\times10^4$. If $V_{\hat{P}_f} \geq 0.05$ when checking the sufficiency of the Monte Carlo population, additional $5\times10^4$ samples are generated and added to $\mathbf{S}_{MC}$. To ensure the reliability of the results, 50 runs are executed for each case. For the two cases, the statistical results of 50 runs are summarized in **Table 7** and **Table 8**.

Under the setting of $k = 3.4$, the system's failure probability is about $4.30\times10^{-4}$. As demonstrated in **Table 7**, both AK-MCS+U and the CFAK-MCSs attain high accuracy, with the average relative errors in failure probabilities being within 0.3%. The quantity of DoE samples required to construct the surrogate model in CFAK-MCS$^b$ exceeds than that of AK-MCS+U. This can be ascribed to the optimization algorithm's capacity to identify points with a lower objective function value in the solution space, thereby delaying the surrogate model construction. In terms of CPU time, CFAK-MCS$^b$ consumes significantly less time than AK-MCS+U due to the difference in the number of LSF predictions. In each run, the number of LSF predictions in AK-MCS+U is approximately $1.460\times10^7$ while that in CFAK-MCS$^b$ is only about $1.75\times10^5$, resulting in a significantly reduced time consumption for CFAK-MCS$^b$. By employing the CFAK-MCS$^p$ and CFAK-MCS$^c$, the number of DoE samples required for surrogate model construction can be further diminished to be less than that of AK-MCS+U while maintaining an acceptable relative error in failure probability.

**Table 7** Results of modified system for case of $k = 3.4$ (50 runs).

| Method | $\hat{P}_f$ /$10^{-4}$ | $\sigma_{\hat{P}_f}$ /$10^{-6}$ | $\bar{\varepsilon}_{\hat{P}_f}$ /% | $\bar{N}_G$ | $\bar{T}$ /s | $\bar{N}_{pred}$ |
|---|---|---|---|---|---|---|
| MCS | 4.2994 | 2.0674 | - | $1\times10^6$ | - | - |
| AK-MCS+U | 4.2926 | 2.0767 | 0.211 | 38.55 | 86.73 | $1.460\times10^7$ |
| CFAK-MCS$^b$ | 4.2970 | 2.0913 | 0.094 | 44.60 | 2.49 | $1.980\times10^5$ |
| CFAK-MCS$^p$ | 4.2988 | 2.0634 | 0.032 | 38.00 | 2.16 | $1.650\times10^5$ |
| CFAK-MCS$^c$ | 4.3010 | 2.0547 | 0.074 | 36.70 | 2.24 | $1.585\times10^5$ |

**Table 8** Results of modified system for case of $k = 3.9$ (50 runs).

| Method | $\hat{P}_f$ /$10^{-5}$ | $\sigma_{\hat{P}_f}$ /$10^{-7}$ | $\bar{\varepsilon}_{\hat{P}_f}$ /% | $\bar{N}_G$ | $\bar{T}$ /s | $\bar{N}_{pred}$ |
|---|---|---|---|---|---|---|
| MCS | 5.7718 | 3.1146 | - | $8\times10^6$ | - | - |
| AK-MCS+U | 5.7653 | 3.1684 | 0.226 | 42.03 | 6785.64 | $5.004\times10^8$ |
| CFAK-MCS$^b$ | 5.7713 | 3.1238 | 0.095 | 40.32 | 17.45 | $1.766\times10^5$ |
| CFAK-MCS$^p$ | 5.7688 | 3.1533 | 0.134 | 31.88 | 14.90 | $1.344\times10^5$ |
| CFAK-MCS$^c$ | 5.7693 | 3.0934 | 0.139 | 30.64 | 14.80 | $1.297\times10^5$ |

Under the setting of $k = 3.9$, the failure probability is reduced by an order of magnitude compared to that under the setting of $k = 3.4$, significantly increasing the difficulty of reliability analysis. The CFAK-MCSs achieve extremely high surrogate model accuracy, with an average relative error in failure probability of only 0.139% or lower. As indicated by **Table 8**, the number of DoE samples required for the CFAK-MCSs to construct the surrogate model is less than that of AK-MCS+U. To ensure the reliability of estimated failure probability, AK-MCS+U necessitates a large number of candidate samples, leading to a substantial increase in the number of LSF predictions during surrogate model construction and a corresponding increase in CPU time. As indicated in **Table 8**, AK-MCS+U requires approximately $5.004\times10^{8}$ LSF predictions for candidate samples during surrogate model construction, resulting in a CPU time of nearly 2 hours. In contrast, the CFAK-MCSs require only $1.77\times10^{5}$ LSF predictions or fewer during surrogate model construction, resulting in CPU time that is about 1/400 of that of AK-MCS+U. Due to the impact of penalty intensity control and density control, CFAK-MCS$^{c}$ demonstrates highest efficiency among the CFAK-MCSs.

For enhanced clarity, the predicted LSE and distribution of DoE samples derived from one of the 50 runs by CFAK-MCS$^{b}$ and CFAK-MCS$^{c}$ are shown in **Fig. 15** and **Fig. 16**, respectively. In this run, both CFAK-MCS$^{b}$ and CFAK-MCS$^{c}$ achieve a satisfactory level of solution accuracy for the failure probability, requiring 40 and 29 DoE samples, respectively. **Fig. 15** depicts the predicted LSE with 13 and 40 DoE samples, respectively, and **Fig. 16**, depicts the predicted LSE with 6, 13, 18 and 29 DoE samples. An examination of **Fig. 15** and **Fig. 16** reveals that CFAK-MCS$^{b}$ selects representative sample points proximate to the LSE at the beginning of the surrogate model construction, while the representative sample points obtained by CFAK-MCS$^{c}$ at the start of surrogate model construction progressively deviate from the sampling center, indicating the influence of penalty intensity control. It is observable that as the quantity of DoE samples increases, the predicted LSE of the surrogate model incrementally converges towards the true LSE. Ultimately, failure points can be accurately incorporated into the predicted failure regions.

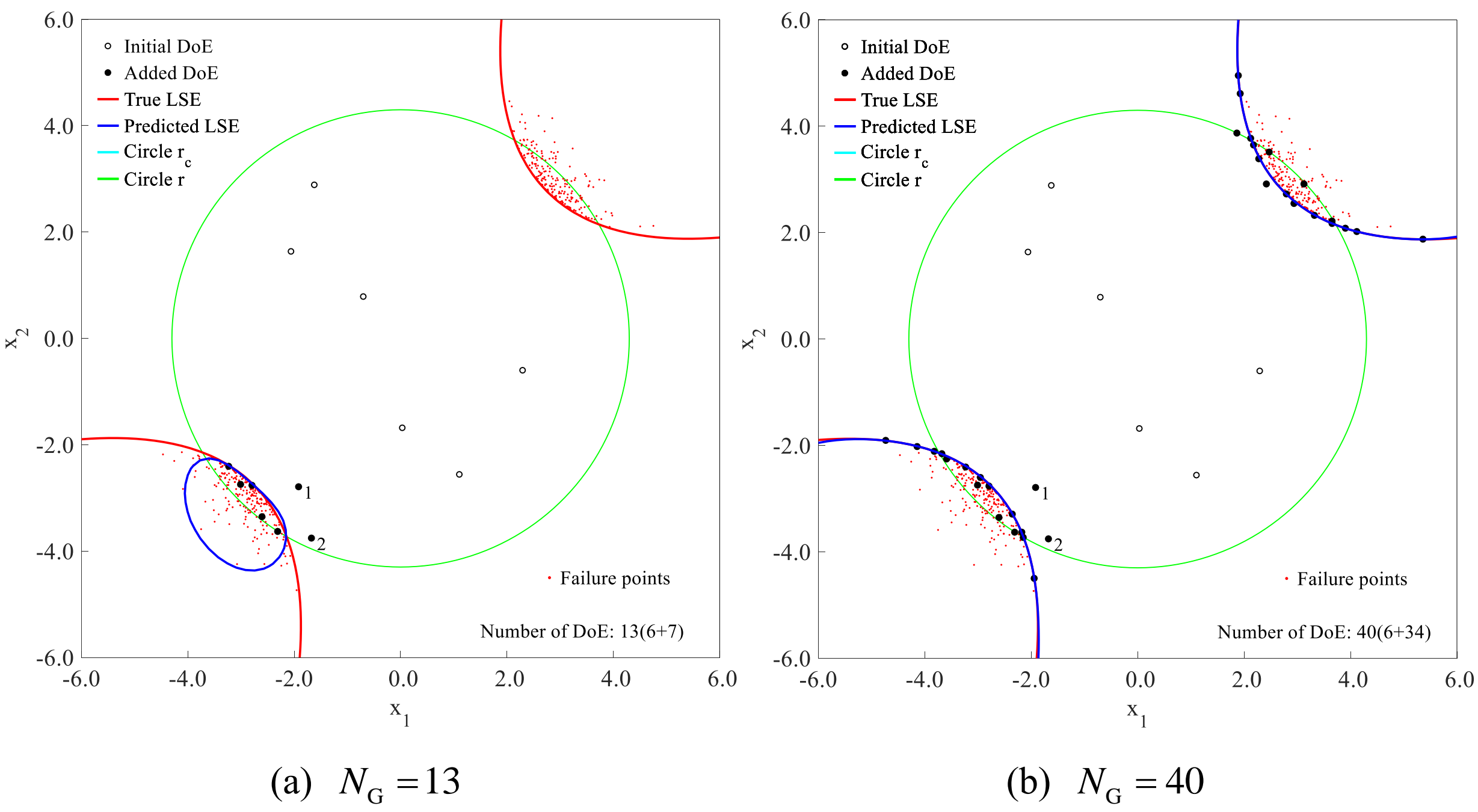

(a) $N_{\mathrm{G}}=13$ (b) $N_{\mathrm{G}}=40$

**Fig. 15**. Predicted LSE and the DoE of CFAK-MCS$^{b}$ ($k=3.9$).

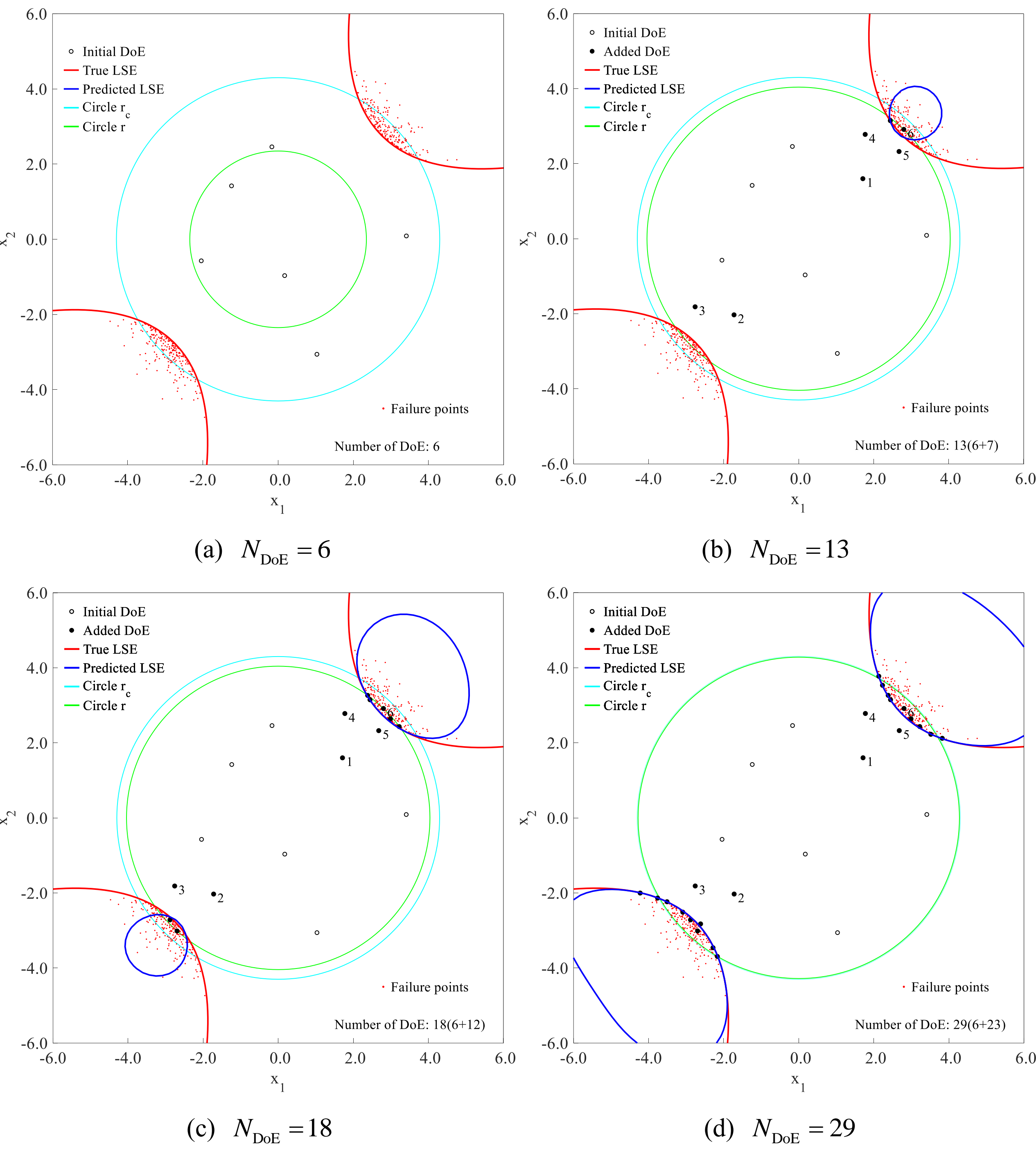

**Fig. 16**. Predicted LSE and the DoE of CFAK-MCS$^{c}$ ($k = 3.9$).

The above numerical experiments demonstrate the superiority of the PSO-based mode used in representative sample selection for problems with small failure probability. In addition, the beneficial impacts of both penalty intensity control and density control are verified again.

*4.2. Study of benchmark problems*

In this section, the computational performance of CFAK-MCS$^c$ is further substantiated through a series of benchmark problems. The benchmark problems include: (1) the LSF with two components, (2) the parabolic LSF, (3) the highly nonlinear LSF, (4) the modified Rastrigin function, (5) dynamic response of a nonlinear oscillator, (6) two-dimensional truss structure, and (7) the high-dimensional LSF. To ensure the robustness of the results, the reported outcomes are based on 50 repeated runs under a specified parameter setting.

For CFAK-MCS$^c$, the configuration of the control parameters for the objective function and the number of initial DoE samples ($N_{\text{DoE_init}}$) is listed in **Table 9**. Some special settings are explained as follows. For the 4$^{\text{th}}$ benchmark, given that the modified Rastrigin function exhibits non-convex and scattered failure domains, $r_d = \delta = 0.1$ is set with further explanation provided in **Sec. 4.2.4**. For the 3$^{\text{rd}}$-5$^{\text{th}}$ benchmark problems, different values of $r_c$ are considered to achieve superior computational performance. In the 6$^{\text{th}}$ benchmark problem, $r_c = 5.0$ is set for the case with a higher failure probability, while $r_c = 5.5$ is set for the other case with a lower failure probability. In the 7$^{\text{th}}$ benchmark problem, the setting of $r_c$ is determined according to the number of dimensions. For cases with 20, 40, 60 and 100 dimensions, $r_c$ is set to 7.0, 8.9, 10.2 and 12.5 ($\gamma \approx 2.5$), respectively.

**Table 9** Configuration of control parameters and number of initial DoE samples.

| Benchmark | Reference | $N_D$ | $r_c$ | $r_d$ | $\lambda$ | $\delta$ | $N_{\text{DoE_init}}$ |
|---|---|---|---|---|---|---|---|
| (1) | [1] | 2 | 4.3 | 0.50 | 0.50 | 0.001 | 6 |
| (2) | [2] | 2 | 4.3 | 0.50 | 0.50 | 0.001 | 6 |
| (3) | [18] | 2 | 3.1-4.3 | 0.50 | 0.50 | 0.001 | 8 |
| (4) | [33] | 2 | 2.7-4.3 | 0.10 | 0.50 | 0.100 | 20 |
| (5) | [37] | 6 | 1.6-4.3 | 0.50 | 0.50 | 0.001 | 13 |
| (6) | [39] | 10 | 5.0/5.5 | 0.50 | 0.50 | 0.001 | 20 |
| (7) | [2] | 20/40/60/100 | 7.0/8.9/10.2/12.5 | 0.50 | 0.50 | 0.001 | 20 |

4.2.1. Benchmark 1: The LSF with two components

The LSF of the system with two components is $G(\mathbf{x}) = \min\{G_1(\mathbf{x}), G_2(\mathbf{x})\}$, where $G_1(\mathbf{x})$ and $G_2(\mathbf{x})$ are the LSFs of the two components, defined as [1, 2]

$$\begin{aligned} G_1(\mathbf{x}) &= 3 - x_2 + \exp\left(-x_1^2 / 10\right) + \left(x_1 / 5\right)^4 \\ G_2(\mathbf{x}) &= 8 - x_1 x_2 \end{aligned} \tag{31}$$

in which $x_1$ and $x_2$ are independent standard normal distributed random variables. The LSF $G_1(\mathbf{x})$ features a single MPP at (0, 4) and the LSF $G_2(\mathbf{x})$ features two MPPs at $\left(2\sqrt{2},2\sqrt{2}\right)$ and $\left(-2\sqrt{2},-2\sqrt{2}\right)$.

In the present work, AK-MCS+U, AWL-MCS and CFAK-MCS$^c$ are conducted to predict the failure probability. The size of Monte Carlo population for failure probability estimation is set to $5\times10^6$. For AK-MCS+U, the number of incremental samples for CSP considering the sufficiency is set to $5\times10^4$. The results of the three surrogate model methods and the results provided by Tabandeh et al. [2], including MCS and three IS methods based on their standard implementations, are listed in **Table 10**. For a more intuitive understanding, the predicted LSE and the DoE samples for one of the 50 runs using CFAK-MCS$^c$ are shown in **Fig. 17**. As indicated by **Table 10**, CFAK-MCS$^c$ exhibits superior solution accuracy and efficiency compared to AK-MCS+U. For this benchmark, the number of calls to LSF required by AWL-MCS is significantly smaller than the other methods, because two Kriging models are employed in AWL-MCS to respectively approximate $G_1(\mathbf{x})$ and $G_2(\mathbf{x})$, significantly reducing the complexity of the LSFs that need to be approximated. However, the CPU time needed by AWL-MCS is more than 100 seconds, which is approximately three times that of CFAK-MCS$^c$. Inspired by AWL-MCS, the implementation of CFAK-MCS$^c$ using multiple surrogate models can be considered as a follow-up research work.

**Table 10** Results of the LSF with two components.

| Method | $\hat{P}_f$ /10$^{-5}$ | $\sigma_{\hat{P}_f}$ /10$^{-7}$ | $\varepsilon_{\hat{P}_f}$ /% | $\bar{N}_{\mathrm{G}}$ ($N_{\mathrm{G}}$) | $\bar{T}$ /s |
|---|---|---|---|---|---|
| MCS [2] | 8.70 | - | - | - | - |
| IS (Gaussian mixture) [2] | 8.97 | - | - | 17932 | - |
| IS (Kernel density) [2] | 8.61 | - | - | 9900 | - |
| IS (Gaussian process) [2] | 8.91 | - | - | 148 | - |
| AK-MCS+U | 8.90 | 3.96 | 2.30 | 162.24 | 4657.32 |
| AWL-MCS | 8.87 | 4.46 | 1.95 | 23.20 | 108.45 |
| CFAK-MCS$^c$ | 8.83 | 4.05 | 1.49 | 81.62 | 35.42 |

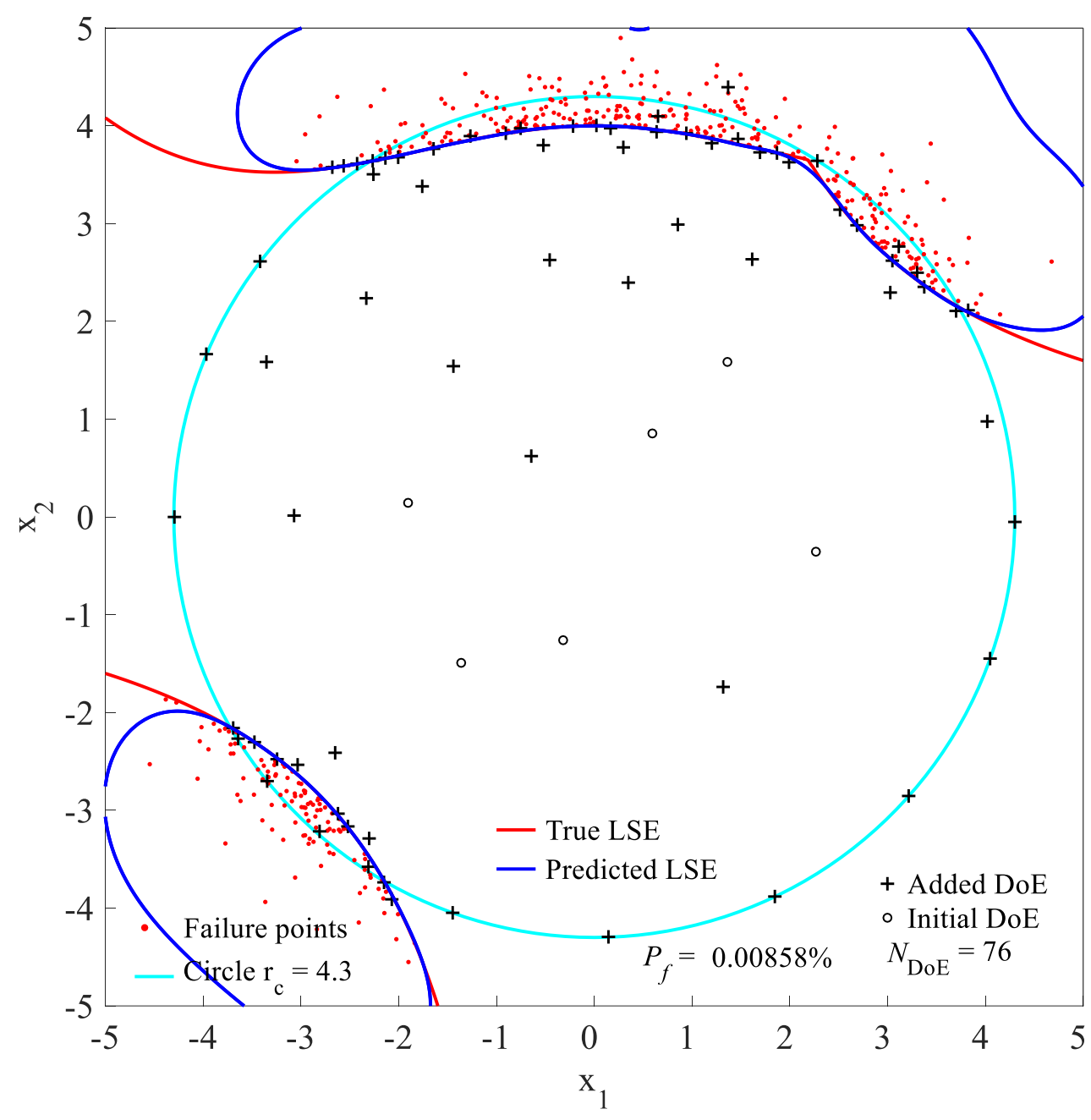

**Fig. 17**. Predicted LSE and DoE for the LSF with two components (CFAK-MCS[c]).

4.2.2. Benchmark 2: The parabolic LSF

The parabolic LSF with two MPPs is defined as [2]

$$G(\mathbf{x}) = 5 - x_2 - 0.5(x_1 - 0.1)^2 \tag{32}$$

where $x_1$ and $x_2$ are independent standard normal distributed random variables. AK-MCS+U, AWL-MCS and CFAK-MCS[c] are conducted to predict the failure probability. The size of Monte Carlo population for failure probability estimation is set to $1\times10^6$. For AK-MCS+U, the number of incremental samples for CSP considering the sufficiency is set to $5\times10^4$. The results of the three surrogate model methods and the estimated failure probability obtained by Tabandeh et al. [2] using MCS, which is taken as reference, are listed in **Table 11**. To be more intuitive, the predicted LSE and the DoE samples for one of the 50 runs using CFAK-MCS[c] are shown in **Fig. 18**. The results indicate that CFAK-MCS[c] can work out high-accurate failure probability in less than 3 seconds of CPU time, which is significantly superior to AK-MCS+U. For this benchmark problem with relatively simple LSF, AWL-MCS still exhibits high computational performance, where only about 13 calls of LSF are required to obtain ideal results. Note that the CPU time of AWL-MCS is slightly higher than that of CFAK-MCS[c].

**Table 11** Results of the parabolic LSF.

| Method | $\hat{P}_f$ /10$^{-3}$ | $\sigma_{\hat{P}_f}$ /10$^{-5}$ | $\varepsilon_{\hat{P}_f}$ /% | $\bar{N}_{\text{G}}$ | $\bar{T}$ /s |
|---|---|---|---|---|---|
| MCS [2] | 3.02 | - | - | - | - |
| AK-MCS+U | 3.02 | 5.26 | 0.0 | 48.78 | 15.63 |
| AWL-MCS | 3.02 | 5.75 | 0.0 | 12.44 | 3.25 |
| CFAK-MCS$^c$ | 3.02 | 4.67 | 0.0 | 20.12 | 2.56 |

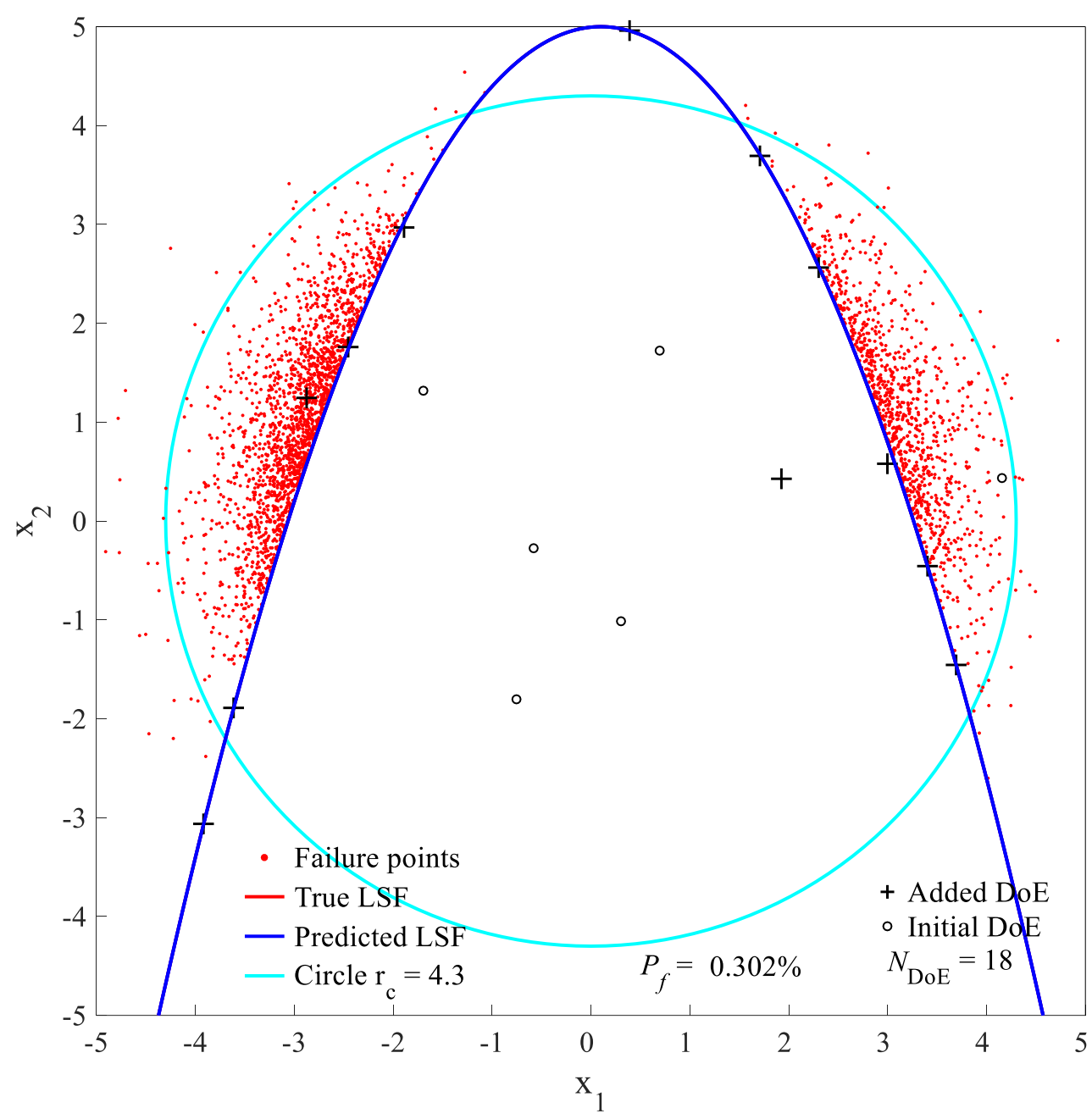

**Fig. 18**. Predicted LSE and DoE for the parabolic LSF (CFAK-MCS$^c$).

4.2.3. Benchmark 3: The highly nonlinear LSF

A 2D analytical example previously studied in [18, 50, 51] is chosen to test the performance of CFAK-MCS$^c$ on a high nonlinearity problem with a single failure region. The LSF is given as [18]

$$G(\mathbf{x}) = 1.2 - \frac{1}{20}\left(x_1^2 + 4\right)\left(x_2 - 1\right) + \sin\left(\frac{5}{2}x_1\right) \tag{33}$$

where $x_1$ and $x_2$ are two independent standard normal distributed random variables. The outcomes of CFAK-MCS$^c$ are compared with those obtained by MCS, FORM, SORM, REIF [18], AK-MCS [16], the method with Error-based Stopping Criterion (ESC) [52], AWL-MCS [40] and the Adaptive Kriging method combining Sampling region scheme and Error-based stopping criterion (AKSE) [37]. The failure probability estimated by MCS with $1\times10^6$ samples is used as the reference. For the second stage of CFAK-MCS$^c$, the estimated failure probability is obtained by predicting $1\times10^6$ samples. Different settings of $r_c$ are considered for CFAK-MCS$^c$ and all of the results obtained by different methods are summarized in **Table 12**, where $\hat{\beta}$ represents the estimated reliability index determined according to the estimated failure probability $\hat{P}_f$. Note that the results of MCS, FORM, SORM, REIF, AK-MCS+U/EFF, ESC+U and AKSE/AKSE-b in **Table 12** are provided by Wang et al. [37]. Additionally, **Fig. 19** shows the predicted LSE, DoE, and estimated failure probabilities for one of the 50 runs using CFAK-MCS$^c$ under different settings of $r_c$.

**Table 12** Results of the highly nonlinear LSF.

| Method | $r_c$ | $\hat{P}_f$ /$10^{-3}$ | $\hat{\beta}$ | $\bar{N}_G$ | $\varepsilon_{\hat{P}_f}$ /% | $V_{\hat{P}_f}$ /% | $\bar{T}$ /s |
|---|---|---|---|---|---|---|---|
| MCS | - | 4.710 | 2.596 | $1\times10^6$ | - | 1.45 | - |
| FORM | - | 2.563 | 1.949 | 779 | 45.58 | - | - |
| SORM | - | 3.671 | 2.681 | 791 | 22.06 | - | - |
| REIF | - | 4.720 | 2.586 | 42.3 | 0.23 | 0.52 | - |
| AK-MCS+U | - | 4.711 | 2.596 | 58.9 | 0.02 | 2.06 | 30.5 |
| AK-MCS+EFF | - | 4.716 | 2.696 | 57.1 | 0.13 | 2.05 | 29.3 |
| ESC+U | - | 4.708 | 2.597 | 45.3 | 0.04 | 2.06 | 89.6 |
| AKSE | - | 4.714 | 2.596 | 30.7 | 0.08 | 2.05 | 170.3 |
| AKSE-b | - | 4.703 | 2.597 | 30.4 | 0.15 | 2.06 | 3500.4 |
| AWL-MCS | - | 4.687 | 2.598 | 60.8 | 0.49 | 1.46 | 7.61 |
| CFAK-MCS$^c$ | 4.3 | 4.709 | 2.596 | 82.0 | 0.01 | 1.45 | 6.99 |
| | 3.6 | 4.695 | 2.598 | 58.2 | 0.32 | 1.46 | 4.31 |
| | 3.2 | 4.683 | 2.598 | 51.4 | 0.56 | 1.46 | 3.73 |
| | 3.1 | 4.652 | 2.601 | 46.7 | 1.23 | 1.46 | 3.33 |

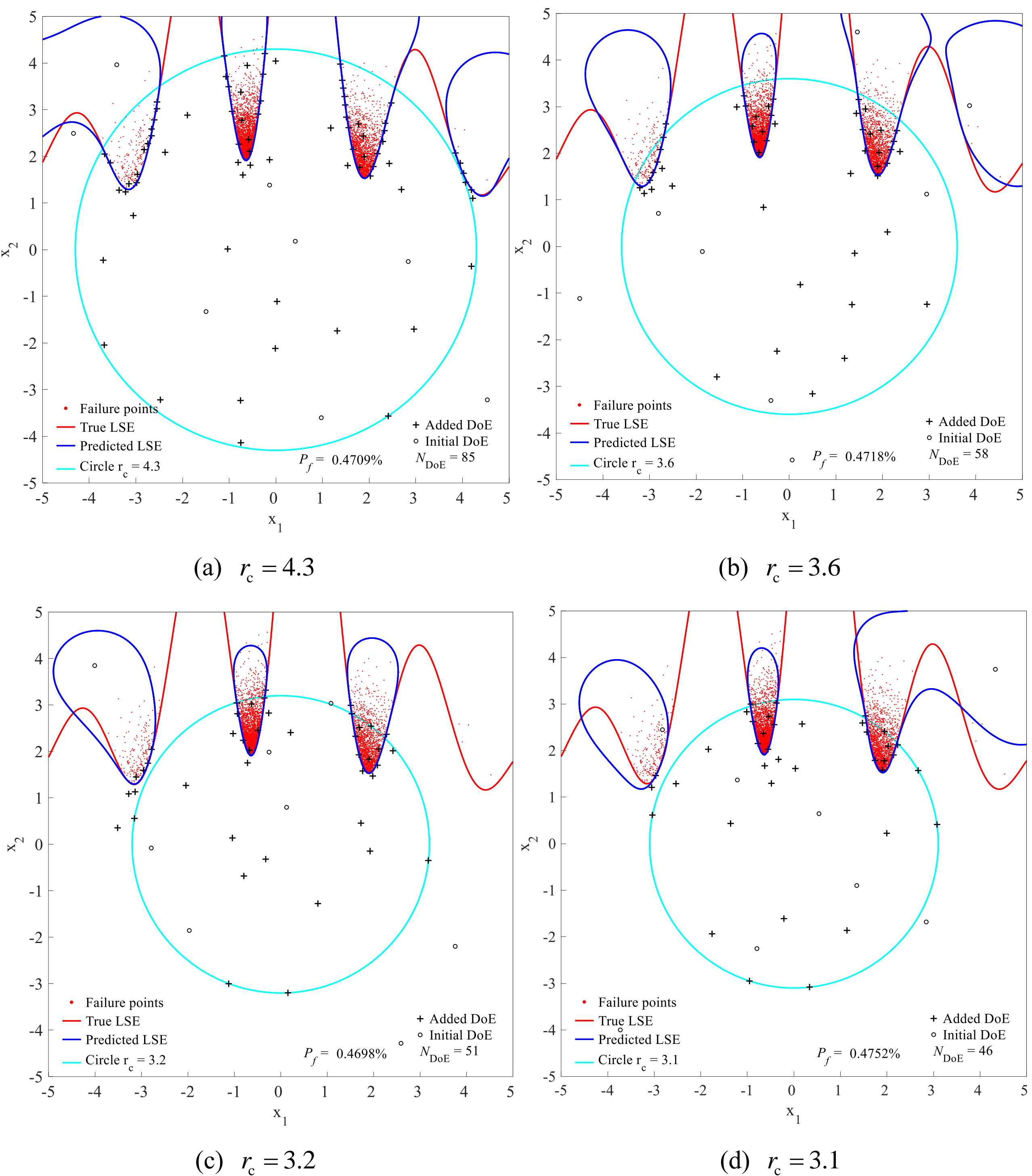

(a) $r_c = 4.3$ (b) $r_c = 3.6$

(c) $r_c = 3.2$ (d) $r_c = 3.1$

**Fig. 19**. Predicted LSE and DoE for the highly nonlinear LSF (CFAK-MCS$^c$).

As indicated in **Table 12**, the failure probabilities estimated by FORM and SORM are relatively inaccurate, exhibiting substantial errors despite necessitating a larger number of calls to LSF compared to the surrogate-based methods. CFAK-MCS$^c$ achieves acceptable results for this example. When $r_c$ is set to a higher value (such as $r_c = 4.3$), the solution accuracy is high, and the number of DoE samples is large, as depicted in **Fig. 19**a). As the parameter $r_c$ decreases, both the number of DoE samples and the solution accuracy decrease correspondingly. Under the setting of $r_c = 3.6$, the number of DoE samples required to construct the surrogate model is equivalent to that of AK-MCS. Under the setting of $r_c = 3.2$, CFAK-MCS$^c$ achieves an optimal balance between solution accuracy and efficiency. In this setting, the number of DoE samples is fewer than that required by AK-MCS, and the relative error of the estimated failure probability is less than 1.0%. In terms of CPU time, CFAK-MCS$^c$ exhibits a significant advantage. Compared with AK-MCS, the CPU time of CFAK-MCS$^c$ is less than 1/6 of the former when the number of DOE samples required by the two methods is similar. For AWL-MCS, although ideal results as listed in **Table 12** can be obtained, a larger size of initial DoE samples ($N_{\text{DoE_init}} = 50$) is required because it is difficult to maintain stable computational performance under the setting of $N_{\text{DoE_init}} = 8$.

4.2.4. Benchmark 4: The modified Rastrigin function

The modified version of the Rastrigin function, which is a highly nonlinear LSF involving non-convex and scattered gaps of failure domains, is expressed as [33, 35, 37]

$$G(\mathbf{x}) = 10 - \sum_{i=1}^{2} \left( x_i^2 - 5\cos\left(2\pi x_i\right) \right) \tag{34}$$

where $x_1$ and $x_2$ are two independent standard normal distributed random variables. The outcomes of CFAK-MCS$^c$ are compared with those of various other methods, among which the estimated failure probability of MCS taken from [53] is used as the reference. The results of other methods including AK-MCS [16], the Reliability method through Error rate-based Adaptive Kriging (REAK) [53], the method combining AK-IS with the Meta-IS (MetaAK-IS$^2$) [54], AWL-MCS [40] and AKSE [37] are taken from the corresponding references. In the second stage of CFAK-MCS$^c$, $6\times10^4$ samples are evaluated to estimate the failure probability. For a LSF that is highly nonlinear and has non-convex and scattered failure domains, it is necessary to relax the distance constraint on the DoE in order to describe the failure regions more accurately. Consequently, the parameter used to describe the distance constraint is set to $r_d = 0.1$. To circumvent difficulties in terminating surrogate model construction due to repeated acquisition of points that are extremely close to the LSE, $\delta$ in Eq. (23) is set to 0.1. This adjustment allows for the identification of representative sample points within the failure regions rather than on the LSE. In addition, the surrogate model is imbued with the capacity to predict the gradient values of LSF in the vicinity of limit state more accurately, which is beneficial for surrogate model construction. Various settings of $r_c$ are considered for CFAK-MCS$^c$, and the results obtained by different methods are summarized in **Table 13**. Meanwhile, **Fig. 20** shows the predicted LSE, DoE, and estimated failure probability of CFAK-MCS[c] under different settings of $r_c$.

**Table 13** Results of the modified Rastrigin function.

| Method | $r_c$ | $\hat{P}_f$ /10^-2 | $\hat{\beta}$ | $\bar{N}_G$ | $\varepsilon_{\hat{P}_f}$ /% | $V_{\hat{P}_f}$ /% | $\bar{T}$ /s |
|---|---|---|---|---|---|---|---|
| MCS | - | 7.308 | 1.453 | $6\times10^4$ | - | 1.45 | - |
| AK-MCS+U | - | 7.340 | 1.451 | 416 | 0.44 | 1.45 | - |
| AK-MCS+EFF | - | 7.340 | 1.451 | 417 | 0.44 | 1.45 | - |
| REAK | - | 7.277 | 1.456 | 401 | 0.42 | <1.50 | - |
| MetaAK-IS$^2$ | - | 7.350 | 1.450 | 480 | 0.57 | 2.50 | - |
| AKSE | - | 7.254 | 1.457 | 263.5 | 0.74 | 1.13 | 805.4 |
| AKSE-b | - | 7.229 | 1.459 | 258.1 | 1.08 | 1.13 | 27593.8 |
| AWL-MCS | - | 7.273 | 1.456 | 299.6 | 0.48 | 1.45 | 18.8 |
| CFAK-MCS[c] | 4.3 | 7.308 | 1.453 | 556.0 | 0.00 | 1.45 | 146.6 |
| | 3.6 | 7.294 | 1.454 | 415.5 | 0.20 | 1.46 | 65.0 |
| | 3.0 | 7.312 | 1.453 | 297.1 | 0.05 | 1.45 | 28.1 |
| | 2.9 | 7.299 | 1.454 | 286.5 | 0.13 | 1.45 | 25.8 |
| | 2.8 | 7.275 | 1.456 | 266.6 | 0.45 | 1.46 | 20.6 |
| | 2.7 | 7.140 | 1.466 | 243.8 | 2.30 | 1.47 | 16.9 |

As shown in **Table 13**, CFAK-MCS[c] achieves high computational accuracy at $r_c \in [2.8, 4.3]$. With the increase of $r_c$, more DoE samples are required to complete the construction of a surrogate model, and the precision of the established surrogate model improves correspondingly. Under the setting of $r_c = 4.3$, the construction of a surrogate model requires approximately 556 DOE samples, and the predicted LSE is almost identical to the true LSE, as shown in **Fig. 20**a). With the decrease of $r_c$, the size of DoE gradually decreases. When the parameter $r_c$ is set to 3.0, the number of DoE samples required is approximately 297, which is fewer than most other methods. As depicted in **Fig. 20**, with the decrease in $r_c$, there are some discrepancies between the predicted LSE and the true LSE in the regions outside the circular area with radius $r_c$. This, to some extent, affects the accuracy of the estimated failure probability obtained by CFAK-MCS[c]. In this benchmark problem, the accuracy of the estimated failure probability hinges on accurately describing the LSE in the area closer to the sampling center. Therefore, even if the LSE cannot be accurately described at the sampling edge, its impact on the accuracy of failure probability estimation is relatively insignificant. It should also be noted that the value of $r_c$ should not be set too low; otherwise, it will lead to a deviation in predicted LSE within large regions and affect failure probability accuracy. Since the proposed method does not require failure probability estimation during surrogate model construction, repeated prediction of sampling populations can be avoided. In fact, sample prediction required in PSO process is far less than that for the CSP. Therefore, the workload is relatively low. As shown in **Table 13**, under the settings of $r_c = 4.3$ and $r_c = 3.0$, the CPU time is only about 147 s and 28 s, respectively, reflecting the efficiency advantage of CFAK-MCS[c]. For this benchmark, a large number of initial DoE samples are required by AWL-MCS to maintain stable computational performance. In fact, the results of AWL-MCS listed in **Table 13** are obtained under the setting of $N_{\text{DoE_init}} = 200$.

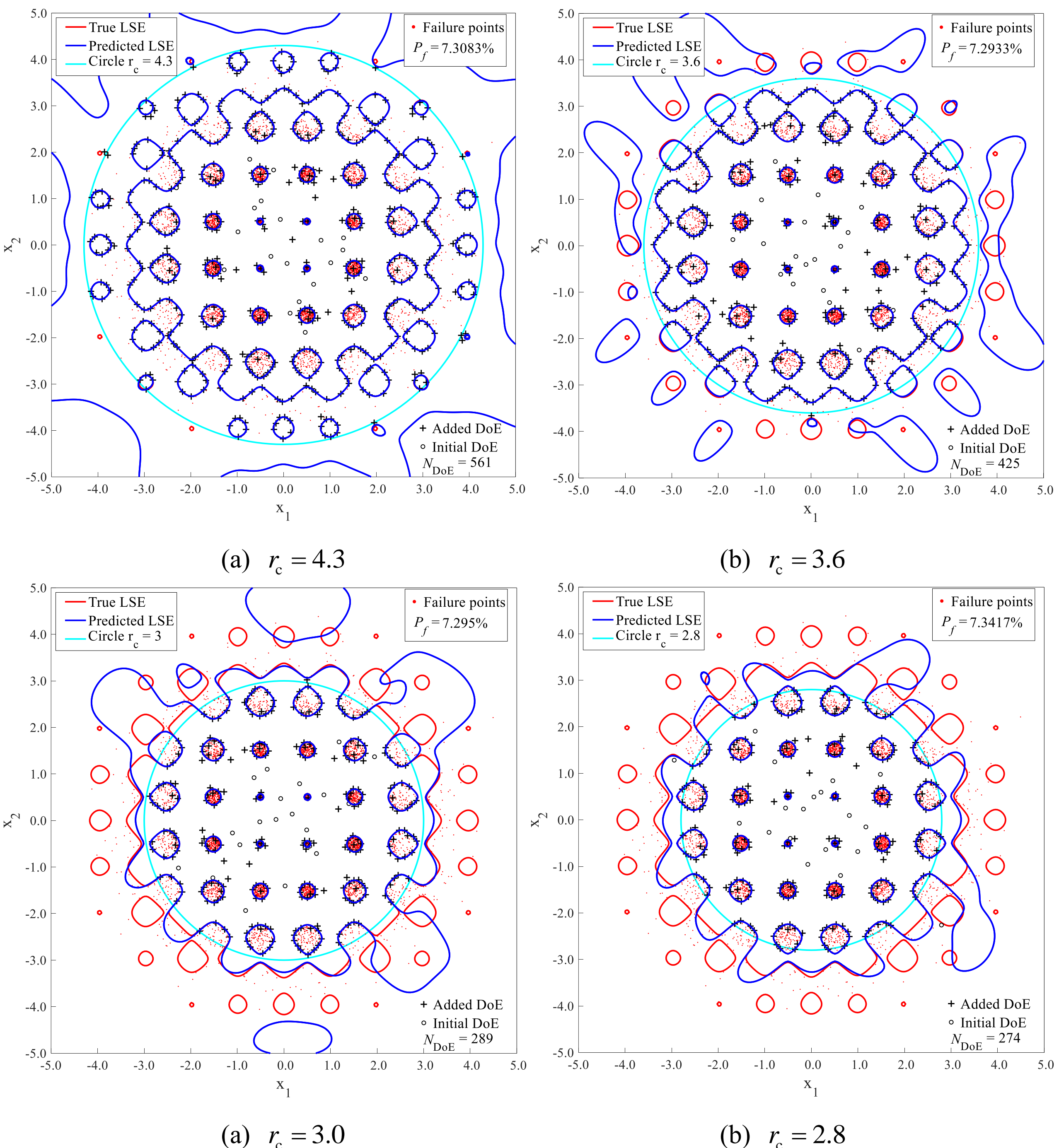

(a) $r_c = 4.3$ (b) $r_c = 3.6$

(a) $r_c = 3.0$ (b) $r_c = 2.8$

**Fig. 20**. Predicted LSE and DoE for the modified Rastrigin function.

4.2.5. Benchmark 5: Dynamic response of a nonlinear oscillator

As shown in **Fig. 21**, the reliability analysis of an undamped single degree of freedom system (a nonlinear oscillator subjected to a rectangular load pulse) is conducted. The LSF of this nonlinear dynamic system is given as [37]

$$G\left(c_{s1},c_{s2},m_s,r_s,t_1,F_1\right)=3r_s-\left|\frac{2F_1}{m_s\omega_0^2}\sin\left(\frac{\omega_0 t_1}{2}\right)\right| \tag{35}$$

where

$$\omega_0=\sqrt{\left(c_{s1}+c_{s2}\right)/m_s} \tag{36}$$

Two cases with different distribution parameters of $F_1$ are analyzed. The distribution parameters of these random variables are listed in **Table 14**.

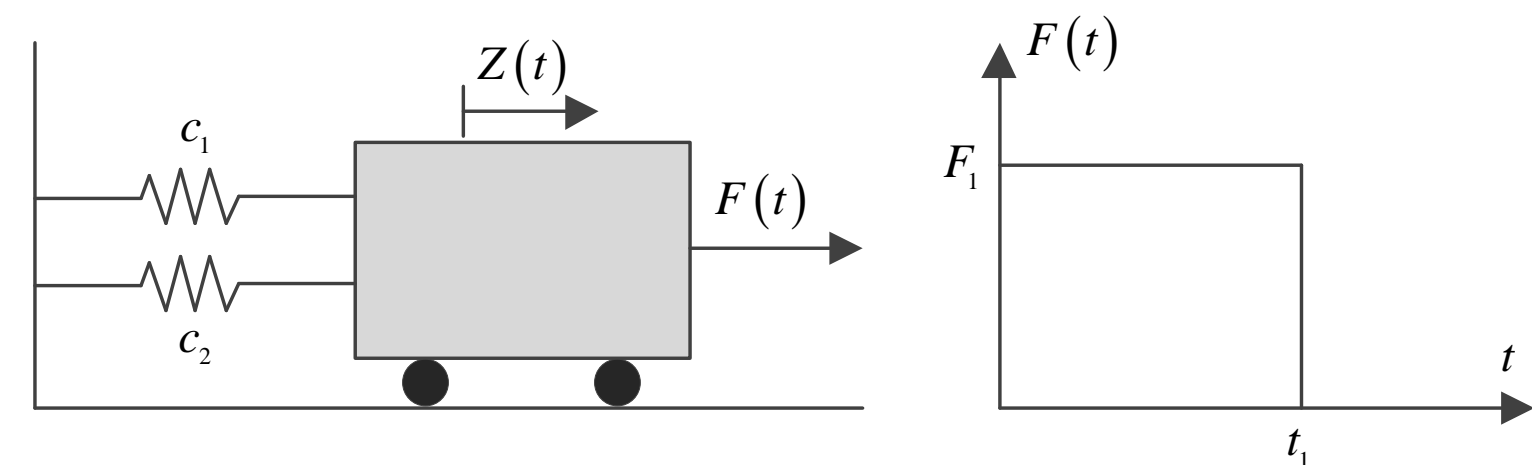

**Fig. 21**. Nonlinear oscillator subjected to a rectangular load pulse.

For Case 1, the results of CFAK-MCS$^c$ are compared with those obtained by MCS, the Active learning method combining Kriging and Subset Simulation (AK-SS) [55]、AK-MSS [33]、AWL-MSS [40]、AK-MCS [16]、ESC [52] and AKSE [37]. The outcomes from these methods are summarized in **Table 15**. Note that the results of MCS, AK-SS, AK-MSS, AK-MCS, ESC and AKSE in **Table 15** are provided by Wang et al. [37]. The estimated failure probability obtained by MCS with $1\times10^6$ samples is used as the reference. The failure probability estimated in the second stage of CFAK-MCS$^c$ is obtained by predicting $1\times10^6$ samples. **Table 15** shows that highly accurate estimated failure probability ( $\varepsilon_{\hat{P}_f}<0.1\%$ ) can be obtained under the setting of $r_c\in\left[2.1,4.3\right]$, demonstrating the advantage of CFAK-MCS$^c$ compared to most listed methods. Similar to benchmarks 3 and 4, the number of DOE samples decreases with the decrease of $r_c$. For Case 1 with a high failure probability of 2.859%, only 36.8 DOE samples are required by CFAK-MCS$^c$ under the setting of $r_c=2.1$. This is fewer than the other methods listed in **Table 15**. Under this setting, the relative error of estimated failure probability is a mere 0.07%, indicative of a relatively high solution accuracy. In terms of CPU time, CFAK-MCS$^c$ takes less than 6 seconds to complete a failure probability estimation, significantly less than other methods listed in **Table 15**.

**Table 14** Statistical information of the random variables for the nonlinear oscillator.

| Random variable | Distribution | Mean | Standard deviation |
|---|---|---|---|
| $m_s$ | Normal | 1 | 0.05 |
| $c_{s1}$ | Normal | 1 | 0.1 |
| $c_{s2}$ | Normal | 0.1 | 0.01 |
| $r_s$ | Normal | 0.5 | 0.05 |
| $t_1$ | Normal | 1 | 0.2 |
| $F_1\left(\text{Case 1}\right)$ | Normal | 1 | 0.2 |
| $F_1\left(\text{Case 2}\right)$ | Normal | 0.6 | 0.1 |

For Case 2, the results of CFAK-MCS$^c$ are compared with those obtained by MCS, AK-IS, AWL-MCS and AKSE, and the results obtained by different methods are summarized in **Table 16**. Note that the results of MCS, AK-IS and AKSE in **Table 16** are provided by Wang et al. [37]. It should be noted that Case 2 is a small failure probability problem ( $\hat{P}_f = 9.090\times10^{-6}$ ), and hence a large size of Monte Carlo population ($1.8\times10^8$) is set to ensure the reliability of MCS results. For the second stage of CFAK-MCS$^c$, two implementations are considered: one estimates failure probability based on a population with $1.8\times10^8$ samples, and the other estimates failure probability by adding $1.0\times10^7$ samples step by step according to **Sec. 3.2.1**. To distinguish the two implementations, the superscript '*' is used to indicate the latter implementation. As shown by **Table 16**, CFAK-MCS$^c$ completes surrogate model construction with a small number of DOE samples and achieves high solution accuracy. In the column of CPU time, results corresponding to CFAK-MCS$^c$ are expressed as '$T_1$+$T_2$', where $T_1$ and $T_2$ represent the CPU time for surrogate model construction and failure probability estimation using the established surrogate model, respectively. As shown in **Table 16**, it only takes 3.3 seconds to complete the construction of a surrogate model in CFAK-MCS$^c$. In other words, most CPU time is spent on predicting all samples using the established surrogate model. For example, it takes approximately 500 seconds to complete prediction of $1.8\times10^8$ samples. It should be noted that the second stage of CFAK-MCS$^c$ does not change the accuracy of a surrogate model but only provides a numerical representation of its accuracy. By using the other implementation, CFAK-MCS$^{c*}$, which has a sample size of $6\times10^7$ in the second stage, not only yields satisfactory solution accuracy and DOE size, but also exhibits a relatively small total CPU time. In fact, several techniques can be used to further reduce time consumption in the second stage, such as the utilization of parallel computing.

**Table 15** Results of dynamic response of a nonlinear oscillator (Case 1).

| Method | $r_c$ | $\hat{P}_f$ /$10^{-2}$ | $\hat{\beta}$ | $\bar{N}_G$ | $\varepsilon_{\hat{P}_f}$ /% | $V_{\hat{P}_f}$ /% | $\bar{T}$ /s |
|---|---|---|---|---|---|---|---|
| MCS | - | 2.859 | 1.902 | $1\times10^6$ | - | 0.58 | - |
| AK-SS | - | 2.833 | 1.906 | 410 | 0.91 | - | - |
| AK-MSS | - | 2.870 | 1.900 | 86 | 0.38 | - | - |
| AWL-MCS | - | 2.826 | 1.907 | 65 | 1.15 | - | - |
| AK-MCS-U | - | 2.850 | 1.903 | 147.2 | 0.31 | 1.85 | 108.4 |
| AK-MCS+EFF | - | 2.867 | 1.901 | 126.8 | 0.28 | 1.84 | 86.7 |
| ESC+U | - | 2.866 | 1.901 | 56.2 | 0.24 | 1.84 | 42.1 |
| AKSE | - | 2.862 | 1.902 | 38.6 | 0.10 | 1.84 | 36.4 |
| AKSE-b | - | 2.851 | 1.903 | 42.0 | 0.28 | 1.85 | 1371.7 |
| CFAK-MCS[c] | 4.3 | 2.860 | 1.902 | 74.2 | 0.03 | 0.58 | 5.98 |
| | 3.6 | 2.861 | 1.902 | 64.9 | 0.04 | 0.58 | 5.06 |
| | 3.1 | 2.861 | 1.902 | 57.1 | 0.08 | 0.58 | 4.37 |
| | 2.6 | 2.861 | 1.902 | 46.3 | 0.05 | 0.58 | 3.55 |
| | 2.1 | 2.857 | 1.902 | 36.8 | 0.07 | 0.58 | 2.97 |
| | 1.6 | 2.863 | 1.901 | 37.9 | 0.14 | 0.58 | 3.03 |

**Table 16** Results of dynamic response of a nonlinear oscillator (Case 2).

| Method | $r_c$ | $\hat{P}_f$ /$10^{-6}$ | $\hat{\beta}$ | $\bar{N}_G$ | $\varepsilon_{\hat{P}_f}$ /% | $V_{\hat{P}_f}$ /% | $\bar{T}$ /s |
|---|---|---|---|---|---|---|---|
| MCS | - | 9.090 | 4.286 | $1.8\times10^8$ | - | 2.47 | - |
| AK-IS+U | - | 9.108 | 4.286 | 281.6 | 0.20 | 2.94 | 3036.9 |
| AK-IS+EFF | - | 9.161 | 4.284 | 182.3 | 0.78 | 2.59 | 1301.1 |
| AKSE-IS | - | 9.032 | 4.288 | 47.2 | 0.64 | 1.91 | 269.7 |
| AKSE-b-IS | - | 9.130 | 4.285 | 49.3 | 0.44 | 2.92 | 6466.3 |
| AWL-MCS | | 9.126 | 4.285 | 62.6 | 0.41 | 2.47 | 5372.5 |
| CFAK-MCS[c] | 4.3 | 9.114 | 4.286 | 41.4 | 0.26 | 2.47 | 3.3+507.2 |
| | 3.0 | 9.063 | 4.287 | 42.8 | 0.30 | 2.48 | 3.3+516.1 |
| CFAK-MCS[c*] | 4.3 | 9.061 | 4.286 | 41.2 | 0.32 | 4.28 | 3.3+173.8 |
| | 3.0 | 9.054 | 4.287 | 42.3 | 0.40 | 4.29 | 3.3+176.2 |

4.2.6. Benchmark 6: Two-dimensional truss structure

**Fig. 22** illustrates a truss structure with 23 bar elements and 10 random variables [29, 39]. This structure is employed to ascertain the applicability of CFAK-MCS$^c$ in problems with non-Gaussian distributed random variables and implicit LSF. The 10 independent variables include cross sections and Young's modulus for horizontal and sloping bars, which are denoted as $A_1, E_1$ and $A_2, E_2$ respectively, and the six applied loads denoted as $P_1$ - $P_6$. **Table 17** provides the distribution information for these 10 random variables. The LSF for structural reliability analysis is defined as

$$G(\mathbf{x}) = v_{\max} - |\Delta(\mathbf{x})| \tag{37}$$

where $v_{\max}$ is the maximum limit deflection and $\Delta(\mathbf{x})$ represents the deflection of the midspan in **Fig. 22**, which is determined through finite element program (made in MATLAB) according to the realization of 10 random variables ($\mathbf{x}$). For the cases of $v_{\max} = 0.11$ and $v_{\max} = 0.14$, the reference failure probabilities are approximately $8.87\times10^{-3}$ and $3.45\times10^{-5}$ [56], respectively. Therefore, for the two cases, $r_c = 5.0$ and $r_c = 5.5$ is set, respectively, and the size of Monte Carlo population for failure probability estimation is set to $1\times10^6$ and $2\times10^7$, respectively.

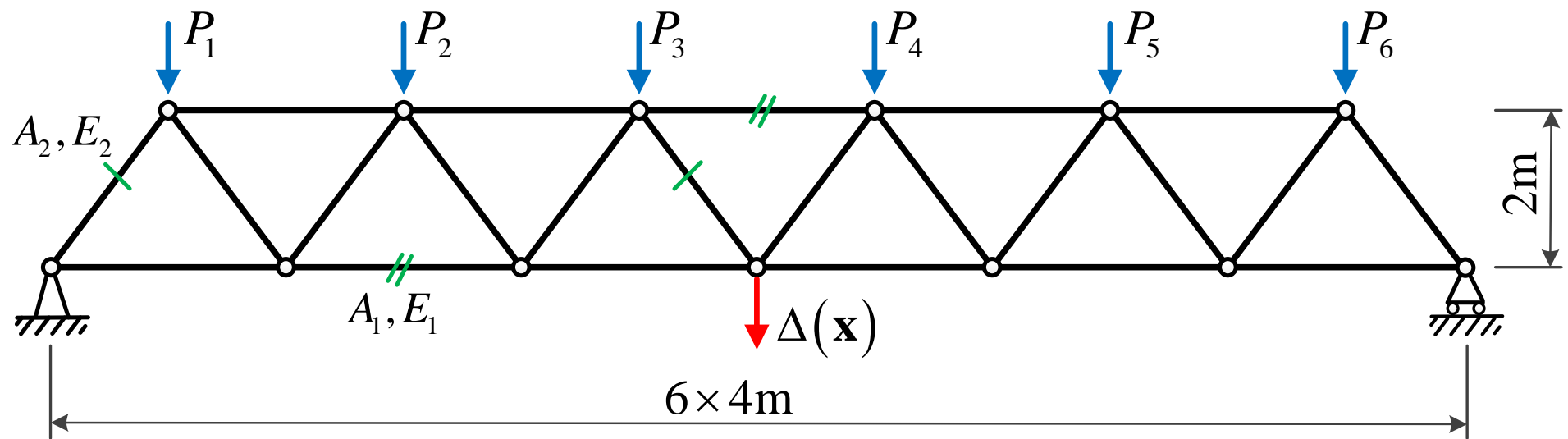

**Fig. 22**. Truss structure with 23 bar elements.

**Table 17** Random variable distribution information for the truss structure.

| Random variable | Distribution | Mean | Standard deviation |
|---|---|---|---|
| $P_1 - P_6$ | Gumbel | $5\times10^4$ | $7.5\times10^3$ |
| $A_1\left(\mathrm{m}^2\right)$ | Lognormal | $2\times10^{-3}$ | $2\times10^{-4}$ |
| $A_2\left(\mathrm{m}^2\right)$ | Lognormal | $1\times10^{-3}$ | $1\times10^{-4}$ |
| $E_1\left(\mathrm{Pa}\right)$ | Lognormal | $2.1\times10^{11}$ | $2.1\times10^{10}$ |
| $E_2\left(\mathrm{Pa}\right)$ | Lognormal | $2.1\times10^{11}$ | $2.1\times10^{10}$ |

In this work, MCS and CFAK-MCS$^c$ are conducted to estimate the failure probabilities of the two situations ($v_{\max} = 0.11$ and $v_{\max} = 0.14$). The results obtained from this study are compared with those derived from IS [56], the method based on Polynomial Chaos Expansions (PCE) [56] and RBF-GA [39]. These comparisons are presented in **Table 18** and **Table 19** for the two situations, respectively. For CFAK-MCS$^c$, the outcomes of the first three runs and the average results, denoted as "average50", are displayed. As indicated by **Table 18** and **Table 19**, the estimated failure probabilities worked out by CFAK-MCS$^c$ are basically consistent with the results of MCS in both situations. Compared with the PCE related methods, CFAK-MCS$^c$ not only reduces the requirement of LSF calls, but also ensures solution accuracy in small failure probability situation. Compared with RBF-GA, CFAK-MCS$^c$ can achieve satisfactory solution accuracy with fewer calls to LSF in the situation with small failure probability, thereby verifying the applicability of CFAK-MCS$^c$ in practical engineering. Specifically, a comparative analysis of the CPU time consumed by both MCS and CFAK-MCS$^c$ is conducted, with the data presented in **Table 18** and **Table 19**. Although the CPU time required for conducting a finite element analysis is only about $4.25\times10^{-4}$ seconds, the time cost required to execute MCS is still quite high. Taking the second situation ($v_{\max}=0.14$) as an example, the CPU time spent on MCS is approximately 2.35 hours. With CFAK-MCS$^c$ used, due to the significant reduction in finite element analysis, it only takes about 125 seconds, while the surrogate model construction takes only 8.5 seconds.

**Table 18** Results of the truss structure with $v_{\max}=0.11$.

| Method | | $\hat{P}_f$ /10$^{-3}$ | $\bar{N}_{\mathrm{G}}$ ($N_{\mathrm{G}}$) | $\varepsilon_{\hat{P}_f}$ /% | $V_{\hat{P}_f}$ /% | $\bar{T}$ /s |
|---|---|---|---|---|---|---|
| IS [56] | | 8.866 | $2\times10^5$ | - | - | - |
| Full PCE [56] | | 8.866 | 443 | 0 | - | - |
| Sparse PCE [56] | | 8.866 | 207 | 0 | - | - |
| MCS [39] | | 8.853 | $1\times10^6$ | - | 1.06 | - |
| RBF-GA [39] | Average50 | 8.900 | 54.2 | 0.53 | 1.06 | - |
| MCS | | 8.875 | $1\times10^6$ | - | 1.06 | 438.7 |
| CFAK-MCS$^c$ | Run 1 | 8.955 | 79 | 0.90 | 1.05 | 2.2+4.7 |
| | Run 2 | 8.868 | 94 | 0.08 | 1.06 | 2.7+5.6 |
| | Run 3 | 9.021 | 100 | 1.39 | 1.05 | 3.0+6.1 |
| | Average50 | 8.923 | 84.9 | 0.54 | 1.05 | 2.4+5.1 |

Note: the failure probabilities of IS, full PCE and sparse PCE are provided by Ref. [39].

**Table 19** Results of the truss structure with $v_{\max} = 0.14$.

| Method | | $\hat{P}_f$ /10^-5 | $\bar{N}_G$ ($N_G$) | $\varepsilon_{\hat{P}_f}$ /% | $V_{\hat{P}_f}$ /% | $\bar{T}$ /s |
|---|---|---|---|---|---|---|
| IS [56] | | 3.446 | 2×10^5 | - | - | - |
| Full PCE [56] | | 2.673 | 443 | 22.43 | - | - |
| Sparse PCE [56] | | 2.351 | 207 | 31.78 | - | - |
| MCS [39] | | 3.364 | 2×10^7 | - | 3.86 | - |
| RBF-GA [39] | Average50 | 3.361 | 112.3 | 0.09 | 3.86 | - |
| MCS | | 3.425 | 2×10^7 | - | 3.82 | 8466.6 |
| CFAK-MCS^c | Run 1 | 3.378 | 132 | 1.37 | 3.85 | 9.9+158.7 |
| | Run 2 | 3.430 | 105 | 0.15 | 3.82 | 8.5+120.3 |
| | Run 3 | 3.451 | 88 | 0.76 | 3.81 | 7.3+101.9 |
| | Average50 | 3.436 | 102.2 | 0.32 | 3.81 | 8.3+116.5 |

Note: the failure probabilities of IS, full PCE and sparse PCE are provided by Ref. [39].

#### 4.2.7. Benchmark 7: The high-dimensional LSF

This section investigates the ability of CFAK-MCS$^c$ in high-dimensional reliability analysis. The high-dimensional LSF is defined as [2]

$$G(\mathbf{x}) = \beta\sqrt{N_D} - \sum_{i=1}^{N_D} x_i \tag{38}$$

where $x_1, x_2, \ldots, x_{N_D}$ are independent standard normal distributed random variables, $N_D$ refers to the problem dimension and $\beta$ is a predetermined parameter of the LSF. It is noteworthy that the exact failure probability is $P_f = \Phi(-\beta)$, regardless of the problem dimension. Tabandeh et al. [2] utilized the LSF to contrast the performances of three IS methods (IS with von Mises-Fisher mixture model, IS with Kernel density model and IS with Gaussian process model) based on their standard implementations, under the setting of $\beta = 3.5$ and $N_D = 20$ where the actual failure probability is $P_{f,ref} = \Phi(-3.5) = 2.32\times10^{-4}$. In their work, the estimated failure probabilities procured by the three IS methods are $2.39\times10^{-4}$, $2.23\times10^{-4}$ and $2.38\times10^{-4}$, respectively, and the required number of LSF calls for the three IS methods is 5755, 13300 and 486, respectively.

In this research, a comparative analysis of the computational performance of AK-MCS+U and CFAK-MCS$^c$ are conducted for the cases of $N_D = 20$, $N_D = 40$, $N_D = 60$ and $N_D = 100$ under the setting of $\beta = 3.5$. The Monte Carlo population for failure probability estimation is generated in advance, with a size of $2\times10^6$. For the high-dimensional problem where the solution space is vast, it becomes challenging for the ordinary Kriging model to eliminate points with $U(\mathbf{x}) < 2.0$ within the solution space, even if the scale of DoE is quite large. This results in difficulty in satisfying the stopping criterion $\min(U(\mathbf{x})) \geq 2.0$ when the algorithm has ability to find the point with the lowest value of $U(\mathbf{x})$ in the solution space. Therefore, the size of CSP for AK-MCS+U is set to $3\times10^5$, which may ignore some points with a low value of $U(\mathbf{x})$ and expedite the construction of surrogate model. For the purpose of comparison, the size of DOE in CFAK-MCS$^c$ is set to align with the mean count of DOE samples necessitated by AK-MCS+U, for the problem with the same dimension. The outcomes of AK-MCS+U and CFAK-MCS$^c$ are presented in **Table 20**.

**Table 20** Results of the high-dimensional LSF.

| $N_D$ | Method | $\hat{P}_f$ /$10^{-4}$ | $\hat{\beta}$ | $\varepsilon_{\hat{P}_f}$ /% | $\bar{N}_{\mathrm{G}}$ | $\bar{T}$ /s | $\bar{N}_{\mathrm{pred}}$ |
|---|---|---|---|---|---|---|---|
| 20 | AK-MCS+U | 2.293 | 3.504 | 1.16 | 74.9 | 89.3 | $1.14\times10^7$ |
| | CFAK-MCS$^c$ | 2.336 | 3.499 | 0.68 | 74 | 22.5 | $2.84\times10^5$ |
| 40 | AK-MCS+U | 2.292 | 3.504 | 1.21 | 131.1 | 452.7 | $2.28\times10^7$ |
| | CFAK-MCS$^c$ | 2.346 | 3.498 | 1.12 | 131 | 85.9 | $5.83\times10^5$ |
| 60 | AK-MCS+U | 2.278 | 3.506 | 1.81 | 220.7 | 2673.2 | $6.70\times10^7$ |
| | CFAK-MCS$^c$ | 2.350 | 3.497 | 1.29 | 220 | 198.6 | $1.06\times10^6$ |
| 100 | AK-MCS+U | 2.241 | 3.510 | 3.41 | 361.2 | 13642.6 | $1.03\times10^8$ |
| | CFAK-MCS$^c$ | 2.279 | 3.506 | 1.77 | 361 | 598.2 | $1.79\times10^6$ |

Note: The reference failure probability is $P_{f,ref} = \Phi(-3.5) = 2.32\times10^{-4}$.

As indicated in **Table 20**, the estimated failure probabilities derived by CFAK-MCS$^c$ are close to the theoretical failure probability, and exhibit higher accuracy compared to the results of AK-MCS+U. Generally, with the increase in dimensionality, the complexity of describing LSE increases, leading to a decrease in solution accuracy. For the problem with $N_D = 100$, the relative error of the estimated failure probability obtained by AK-MCS+U exceeds 3%, while that of CFAK-MCS$^c$ remains within 2%. With the increase in dimensionality, more representative samples are necessitated to describe the LSE. Consequently, the scale of DoE progressively increases as shown in **Table 20**. For the problem with a given dimension, even though the number of DoE samples for the two methods is similar, the efficiency of CFAK-MCS$^c$ is significantly superior to AK-MCS. This is primarily due to the difference in LSF prediction times. CFAK-MCS$^c$ only requires approximately 5000 predictions of LSF for identifying one representative sample point, while AK-MCS requires approximately $3\times10^5$

LSF predictions. As the increase of dimensionality, the time consumption of a single LSF prediction also increases, ultimately increasing the proportion of CPU time between the two methods. Although the CPU time of AK-MCS can be reduced by reducing the size of CSP, the solution accuracy will be difficult to guarantee.

Indeed, high-dimensional problems pose a significant challenge for surrogate model methods, due to the vast solution space. The enhancement of surrogate models and the introduction of a more judicious stopping criterion are pivotal for augmenting the computational efficiency of the surrogate model method. Moreover, the utilization of dimensionality reduction techniques and the amalgamation of diverse surrogate models may potentially yield more efficacious solutions. These propositions represent promising avenues for future exploration in this research domain.

## 5. Conclusions

This study proposes a CSP-free adaptive Kriging surrogate model method for reliability analysis with small failure probabilities. In this method, the surrogate model undergoes iterative refinement, and representative samples are selected for updating the surrogate model through an optimization solution facilitated by the particle swarm optimization algorithm. To achieve an optimal balance between solution accuracy and efficiency, penalty intensity control and density control for the experimental design points are introduced to modify the objective function in optimization. Numerical examples are used to validate the proposed method and evaluate its computational performance. The following conclusions can be drawn:

(1) The incorporation of a particle swarm optimization algorithm effectively enhances the likelihood of obtaining representative samples for the construction of surrogate model. This results in an improvement in solution accuracy for an adaptive Kriging surrogate model method in reliability analysis.

(2) The proposed CSP-free surrogate model method bifurcates the process of surrogate model construction and failure probability estimation into two distinct stages, leading to a more transparent implementation process. This approach eliminates the need for repeated estimation of the failure probability during surrogate model construction. Moreover, this method exhibits the capability to capture representative sample points in regions of sparse distribution. Therefore, the proposed method is particularly suitable for reliability analysis with small failure probabilities.

(3) The introduction of penalty intensity control and density control to the objective function is instrumental in the judicious selection of representative samples for surrogate model updates, thereby improving the computational performance of the adaptive surrogate model methods.

## Acknowledgments

The project is funded by the National Natural Science Foundation of China (Grant No. 52178209, Grant No. 51878299) and Guangdong Basic and Applied Basic Research Foundation, China (Grant No. 2020A1515010611, Grant No. 2021A1515012280).